\documentclass[11pt]{article}
\title{Particle-Like Solutions of the Einstein-Dirac Equations}
\author{Felix Finster\thanks{Research supported by the 
Deutsche Forschungsgemeinschaft and the Schweizerischer 
Nationalfonds.},\ 
Joel Smoller\thanks{Research supported in part by the NSF, Grant No.\ 
DMS-G-9501128.}, and Shing-Tung Yau\thanks{Research supported in part 
by the NSF, Grant No.\ 33-585-7510-2-30.}}
\date{January 1998}

\newcommand{\spc}{\;\;\;\;\;\;\;\;\;\;}
\newcommand{\bra}{\mbox{$< \!\!$ \nolinebreak}}
\newcommand{\ket}{\mbox{\nolinebreak $>$}}

\newcommand{\R}{\mbox{\rm I \hspace{-.8 em} R}}
\newcommand{\1}{\mbox{\rm 1 \hspace{-1.05 em} 1}}

\newcommand{\Tr}{\mbox{Tr\/}}

\begin{document}
\include{epsf}

\maketitle

\begin{abstract}
The coupled Einstein-Dirac equations for a static, spherically 
symmetric system of two fermions in a singlet spinor state are 
derived. Using numerical methods, we construct an infinite number of 
soliton-like solutions of these equations.
The stability of the solutions is analyzed. For weak coupling (i.e., 
small rest mass of the fermions), all the solutions are linearly stable
(with respect to spherically symmetric perturbations), whereas for stronger
coupling, both stable and unstable solutions exist.
For the physical interpretation, we discuss how the energy of the 
fermions and the (ADM) mass behave as functions of the rest mass of 
the fermions. Although gravitation is not 
renormalizable, our solutions of the Einstein-Dirac equations are 
regular and well-behaved even for strong coupling.
\end{abstract}

\section{Introduction}
In recent years, there has been much interest in the coupling 
of Einstein's field equations to Yang-Mills equations.
In this case, the attractive gravitational force is balanced by the 
repulsive Yang-Mills force, and this interaction has led to many 
interesting and surprising results; see for example
\cite{2}-\cite{14}.
In this paper, we consider the coupling of Einstein's equations to the
Dirac equation. Here the necessary repulsive mechanism is provided by the
Heisenberg Uncertainty Principle.

The Einstein-Dirac equations take the form
\begin{equation}
	R^i_j-\frac{1}{2} \:R\:\delta^i_j \;=\; -8 \pi \:T^i_j \;\;\;,\spc
	(G-m) \:\Psi \;=\; 0 \;\;\; ,
	\label{1.1}
\end{equation}
where $T^i_j$ is the energy-momentum tensor of the Dirac particle, 
$G$ denotes the Dirac operator (see \cite{8}), and $\Psi$ is the wave 
function of a fermion of mass $m$.
As in the above-mentioned earlier studies, we consider static, spherically
symmetric solutions. Since the spin of a fermion has an 
intrinsic orientation in space, a system consisting of a single Dirac 
particle cannot be spherically symmetric. In order to maintain 
the spherical symmetry, we are led to the study of two fermions having 
opposite spin; i.e.\ to a singlet spinor state. Of course, such a 
configuration does not represent a realistic physical system due to 
the absence of the electromagnetic interaction. More precisely,
neglecting the electromagnetic interaction corresponds to the
limiting case where the masses of the fermions become so large (of the 
order $({\mbox{Planck length}})^{-1}$) that the gravitational 
interaction becomes the dominant force. Nevertheless, we view this study
as a model problem worth considering in order to get some understanding of
the equations and their solutions.
In a future publication, we will consider the more physically 
realistic situation where the Einstein-Dirac equations are coupled 
to an electromagnetic field (Maxwell's equations).

Our results are based on a certain ansatz, whereby we reduce the 
4-component Dirac spinors to a 2-component spinor system, 
$\Phi=(\alpha, \beta)$ with real functions $\alpha$, $\beta$.
We show numerically that particle-like solutions of this type exist,
both in the ground state, and in the excited states.
For weak coupling, i.e.\ small mass $m$, the different solutions are 
characterized by the ``rotation number'', $n=0,1,2,\ldots$, of the vector
$(\alpha, \beta)$ (we work in standard units $\hbar=c={\cal{G}}=1$).
The solution with $n=0$ is the ground state, and the solutions with 
$n=1,2,\ldots$ describe the excited states.
For small $m$, the solutions are (linearly) stable with respect to spherically 
symmetric perturbations.
However, as $m$ gets large, several states appear for each $n$.
In fact, for every $n$, the mass spectrum (i.e., the plot of the binding energy
vs.\ the rest mass) 
is a  spiral curve which tends to a limiting configuration.
This surprising result shows that for parameter values on this 
limiting configuration, there are an infinite number of excited states
``in the $n^{\mbox{\tiny{th}}}$ mode'', while for parameter values near this 
limiting configuration, there are still a large, but finite number of such
excited states.
Furthermore, using topological methods and bifurcation 
theory (see \cite[part IV]{14a}), we show that in every mode, the 
stable solutions must become unstable as the binding energy increases.
Although gravitation is not renormalizable 
(which means that the problem cannot be treated in a perturbation 
expansion), our solutions of the Einstein-Dirac equations are regular 
and well-behaved even for strong coupling.

\section{The Dirac Operator}
\setcounter{equation}{0}
In this section, we shall derive the form of the Dirac operator in
the presence of a static, spherically symmetric gravitational field.
In preparation, we first give a brief mathematical introduction 
of the Dirac theory in curved space-time.
The Dirac operator $G$ is a differential operator of first order
\begin{equation}
	G \;=\; i G^j(x) \frac{\partial}{\partial x^j} \:+\: B(x) \spc ,
	\label{do}
\end{equation}
where the Dirac matrices $G^j(x)$, $(j=0,1,2,3)$, and $B(x)$ are $(4 
\times 4)$ matrices, which depend on the space-time point $x$.
The Dirac matrices and the Lorentzian metric are related by
\begin{equation}
	g^{jk} \;=\; \frac{1}{2} \: \left\{ G^j,\:G^k \right\} \spc ,
	\label{dm}
\end{equation}
where $\{.,.\}$ is the anti-commutator
\[ \{G^j,\:G^k\} \;=\; (G^j \:G^k \:+\: G^k \:G^j) \spc . \]
The basic difficulty with Dirac spinors in curved space-time is that, for
a given Lorentzian metric, the Dirac matrices are not uniquely determined
by the anti-commutation relations (\ref{dm}). One way of fixing the Dirac
matrices is provided by the formalism of spin and frame bundles (see e.g.\
the first section of \cite{PT}). In this formulation, one chooses a
frame $(u_a)_{a=0,\ldots,3}$ and represents the Dirac matrices as linear
combinations of the Dirac matrices $\gamma_a$ of Minkowski space,
\[ G^j(x) \;=\; \sum_{a=0}^3 u^j_a(x) \:\gamma^a \;\;\; . \]
The matrix $B(x)$ is composed of the so-called spin connection coefficients,
involving first partial derivatives of the metric and of the frame.
It is quite common to choose for $(u_a)$ a Newman-Penrose null frame;
this choice is particularly convenient for metrics of Petrov type D (see
\cite{C} for an introduction to the Newman-Penrose formalism and many
applications, especially in the Kerr background).
More generally, it is shown in \cite{8} that all choices of Dirac matrices
satisfying (\ref{dm}) yield unitarily equivalent Dirac operators. Furthermore,
\cite{8} gives explicit formulas for the matrix $B$ in terms of the Dirac
matrices $G^j$. We prefer working with the formalism of \cite{8} in the
following, because it gives us more flexibility in choosing the Dirac
matrices.

The wave function $\Psi$ of a Dirac particle is a solution of the 
Dirac equation
\begin{equation}
	(G-m) \:\Psi \;=\; 0 \spc .
	\label{2.3a}
\end{equation}
On the wave functions, two different scalar products can be defined. In 
the first, we integrate the wave functions over all of space-time,
\begin{equation}
	\bra \Psi \:|\: \Phi \ket \;=\; \int \overline{\Psi} \Phi 
	\:\sqrt{|g|} \:d^4 x \spc ,
	\label{sp1}
\end{equation}
where $\overline{\Psi}=\Psi^* \left( \begin{array}{cc}
\1 & 0 \\ 0 & -\1 \end{array} \right)$ is the adjoint spinor (whose 
definition does not depend on the gravitational field;
$0$, $\1$ are $(2 \times 2)$ submatrices), and $g$ 
denotes the determinant of the metric $g_{jk}$.
The scalar product (\ref{sp1}) is indefinite, but it 
is nevertheless useful to us because the Dirac operator is Hermitian 
with respect to it. The second scalar product is defined on the solutions 
of the Dirac equation. For this we choose a space-like hypersurface
${\cal{H}}$ together with a (future-directed) normal vector field 
$\nu$, and set
\begin{equation}
	(\Psi \:|\: \Phi) \;=\; \int_{\cal{H}} \overline{\Psi} G^j \Phi \;
	\nu_j \: d\mu \spc ,
	\label{sp2}
\end{equation}
where $d\mu$ is the invariant measure on the hypersurface ${\cal{H}}$, 
induced by the metric $g_{ij}$.
This scalar product is positive definite, and, as a consequence of the 
current conservation (cf.\ \cite{8})
\begin{equation}
	\nabla_j \: \overline{\Psi} G^j \Phi \;=\; 0 \spc ,
	\label{2.5a}
\end{equation}
it is independent of the choice of the hypersurface ${\cal{H}}$.
In direct generalization of the expression 
$\overline{\Psi} \gamma^0 \Psi$ in Minkowski space (see e.g\ \cite{4}),
the integrand $\overline{\Psi} G^j \Psi \:\nu_j$ is interpreted as the
probability density of the particle. This leads us to normalize solutions
of the Dirac equation by requiring
\begin{equation}
	( \Psi \:|\: \Psi) \;=\; 1 \spc .
	\label{nc}
\end{equation}

We now return to the Dirac operator (\ref{do}).
Suppose that a 4-dimensional space-time with metric $g_{ij}$ is given.
According to \cite{8}, we can choose for $G^j$ any $4 \times 4$ matrices
which are Hermitian with respect to the scalar product (\ref{sp1}) and satisfy
(\ref{dm}). The matrix $B(x)$ involves first derivatives of the Dirac matrices
$G^j$, and from \cite{8}, we have the explicit formulas
\begin{equation}
B(x) \;=\; G^j(x) \: E_j(x) \label{bf}
\end{equation}
with
\begin{eqnarray}
E_j &=& \frac{i}{2}\: \rho (\partial_j \rho) \:-\: \frac{i}{16}\: \Tr 
(G^m \:\nabla_j G^n) \: G_m G_n \:+\: \frac{i}{8}\: \Tr (\rho G_j \:
\nabla_m G^m) \:\rho \label{gf} \\
\rho &=& \frac{i}{4!} \:\epsilon_{ijkl} \:G^i \:G^j \:G^k \:G^l
\label{rho}
\end{eqnarray}
($\epsilon_{ijkl}$ is the totally antisymmetric tensor density).

Now we will specify these formulas for the Dirac operator
to static, spherically symmetric
space-times. In polar coordinates $(t, r, \vartheta, \varphi)$, the 
metric can be written as (cf.\ \cite{1,11})
\begin{eqnarray}
	g_{ij} & = & {\mbox{diag }}(\frac{1}{T^2},\: -\frac{1}{A},\: -r^2,\:
	-r^2 \: \sin^2 \vartheta) \label{29a} \\
	g^{ij} & = & {\mbox{diag }}(T^2,\: -A,\: 
	-\frac{1}{r^2},\:
	-\frac{1}{r^2 \: \sin^2 \vartheta} ) \label{29b}
\end{eqnarray}
with volume element
\[ \sqrt{|g|} \;=\; T^{-1} \:A^{-\frac{1}{2}} \:r^2\: |\sin \vartheta|
\spc , \]
where $A=A(r)$ and $T=T(r)$ are positive functions.
We shall use this form of the metric to explicitly calculate the Dirac 
operator (\ref{do}). For the Dirac matrices $G^j(x)$, we take an 
ansatz as a linear combination of the usual $\gamma$-matrices in the
Dirac representation
\begin{equation}
	\gamma^0 \;=\; \left( \begin{array}{cc} \1 & 0 \\ 0 & -\1 
	\end{array} \right) \;\;\;,\spc \gamma^i \;=\;
	\left( \begin{array}{cc} 0 & \sigma^i \\ -\sigma^i & 0
	\end{array} \right) \;\;,\;\;\; i=1,2,3,
	\label{2.3}
\end{equation}
where $\sigma^1, \sigma^2, \sigma^3$ are the Pauli matrices.
In order to satisfy (\ref{dm}), we must transform these Dirac matrices 
of the vacuum into polar coordinates and multiply them by the factors
$T$ and $\sqrt{A}$,
\begin{eqnarray}
G^t &=& T \: \gamma^0 \label{2.13a} \\
G^r &=& \sqrt{A} \:\left( \gamma^1 \:\cos \vartheta \:+\: \gamma^2 \:
	\sin \vartheta \: \cos \varphi \:+\: \gamma^3 \: \sin \vartheta \:
	\sin \varphi \right) \\
G^\vartheta &=& \frac{1}{r} \left( -\gamma^1 \:\sin \vartheta \:+\: \gamma^2 \:
	\cos \vartheta \: \cos \varphi \:+\: \gamma^3 \: \cos \vartheta \:
	\sin \varphi \right) \label{2.13aa} \\
G^\varphi &=& \frac{1}{r \sin \vartheta} \left( -\gamma^2 \: \sin \varphi \:+\:
	\gamma^3 \: \cos \varphi \right) \spc . \label{2.13b}
\end{eqnarray}
This choice is convenient, because it greatly simplifies equations 
(\ref{bf})-(\ref{rho}). Namely, the matrix $\rho$ becomes independent 
of $x$ and coincides with the usual ``pseudo scalar'' matrix 
$\gamma^5$ in the Dirac representation,
\[ \rho \;\equiv\; \gamma^5 \;=\; i 
\:\gamma^0\:\gamma^1\:\gamma^2\:\gamma^3 \;=\; \left( \begin{array}{cc}
0 & \1 \\ \1 & 0 \end{array} \right) \spc . \]
As a consequence, the first and last summands in (\ref{gf}) vanish 
and thus
\begin{eqnarray}
B &=& -\frac{i}{16}\: \Tr 
		(G^m \:\nabla_j G^n) \: G^j \:G_m G_n \nonumber \\
&=& -\frac{i}{16}\: \Tr (G^m \:\nabla_j G^n) \left( \delta^j_m \:G_n 
\:-\: \delta^j_n \:G_m \:+\: G^j \:g_{mn} \:+\: i 
\varepsilon^j_{\;\;mnp} \:\gamma^5 \:G^p \right) \;\;\; . \spc\label{2.16a}
\end{eqnarray}
Using Ricci's Lemma
\[ 0 \;=\; 4\:\nabla_j g^{mn} \;=\; \nabla_j \Tr (G^m \:G^n) \;=\;
\Tr ((\nabla_j G^m) \:G^n) \:+\: \Tr (G^m \:(\nabla_j G^n)) \spc , \]
we conclude that the contributions of the first and second summands in 
the right bracket in (\ref{2.16a}) coincide and that the contribution 
of the third summand vanishes. Using the anti-symmetry of the
$\epsilon$-tensor,
we can, in the contribution of the fourth summand, replace
the covariant derivative by a partial derivative. This gives
\begin{equation}
	B \;=\; \frac{i}{8}\: \Tr (G^n \:\nabla_j G^j) \:G_n \:+\: 
	\frac{1}{16} \:\epsilon^{jmnp} \:\Tr (G_m \:\partial_j G_n) \;
	\gamma^5 G_p \spc .
	\label{2.16b}
\end{equation}
The second summand in (\ref{2.16b}) is zero. Namely, the trace always
vanishes if the tensor indices are all different,
\begin{equation}
	\Tr (G_m \:\partial_j G_n) \;=\; 0 \spc{\mbox{for }}\;\;
	m,j,n=t,r,\vartheta, {\mbox{ or }} \varphi \;\;{\mbox{ and 
	}}\;\; m \neq j \neq n \neq m \;\; ;
	\label{2.16c}
\end{equation}
this can be verified directly using our special ansatz
(\ref{2.13a})-(\ref{2.13b}) for the Dirac matrices. In the first 
summand in (\ref{2.16b}), we can use that $\nabla_j G^j$ is a linear 
combination of the Dirac matrices $G^j$, and thus
\[ \Tr (G^n \:\nabla_j G^j) \:G_n \;=\; 4 \:\nabla_j G^j \spc . \]
We conclude that
\begin{equation}
	B \;=\; \frac{i}{2} \:\nabla_j G^j \spc .
	\label{bf1}
\end{equation}

This form of $B(x)$ as a divergence of the Dirac matrices
allows us to easily check that the Dirac operator is Hermitian with respect to
the scalar product $\bra .\:|\:. \ket$; indeed
\begin{eqnarray*}
\bra G \Psi \:|\: \Phi \ket &=& \int \overline{\left( iG^j 
\frac{\partial}{\partial x^j} + \frac{i}{2} \: \nabla_j G^j \right) \Psi}
\:\Phi \;\sqrt{|g|} \:d^4x \\
&=& \int \overline{\Psi} \left( i G^j \frac{\partial}{\partial x^j}
\:-\: \frac{i}{2} 
\:\nabla_j G^j \right) \Phi \: \sqrt{|g|} \:d^4 x \;+\;
\int \overline{\Psi} \:\left(i \partial_j (\sqrt{|g|} G^j )\right)
\Phi \; d^4x\\
&=& \int \overline{\Psi} \left( i G^j \frac{\partial}{\partial x^j}
\:+\: \frac{i}{2} \:\nabla_j G^j \right) \Phi \: \sqrt{|g|} \:d^4 x
\;=\; \bra \Psi \:|\: G \Phi \ket \spc .
\end{eqnarray*}

In order to calculate the divergence (\ref{bf1}), we first compute
\begin{eqnarray*}
\lefteqn{ \frac{1}{\sqrt{|g|}} \:\partial_t ( \sqrt{|g|} \: G^t ) \;=\; 0 } \\
\lefteqn{ \frac{1}{\sqrt{|g|}} \:\partial_r ( \sqrt{|g|} \: G^r )
\;=\; A^{\frac{1}{2}} \:T \: r^{-2} \: \partial_r
	\left( r^2 \: A^{-\frac{1}{2}} \:T^{-1} \: G^r \right)
\;=\; \left( \frac{2}{r} - \frac{T^\prime}{T} \right) G^r } \\
\lefteqn{ \frac{1}{\sqrt{|g|}} \:\partial_\vartheta ( \sqrt{|g|} \: G^\vartheta ) }\\
 &=& \frac{1}{r \sin \vartheta} \: \partial_\vartheta
	\left( -\gamma^1 \: \sin^2 \vartheta \:+\: \gamma^2 \: \sin \vartheta \:
	\cos \vartheta \: \cos \varphi \:+\: \gamma^3 \: \sin \vartheta \:
	\cos \vartheta \: \sin \varphi \right) \\
&=&\frac{1}{r \sin \vartheta} \left(-2 \gamma^1 \:\sin \vartheta \:
\cos \vartheta \:+\: \gamma^2 \: (\cos^2 \vartheta - \sin^2 
\vartheta) \: \cos \varphi \:+\: \gamma^3 \: (\cos^2 \vartheta - \sin^2 
\vartheta) \: \sin \varphi \right) \\
\lefteqn{ \frac{1}{\sqrt{|g|}} \:\partial_\varphi ( \sqrt{|g|} \: G^\varphi )
\;=\; \frac{1}{r \sin \vartheta} \left( -\gamma^2 \:
	\cos \varphi \:-\: \gamma^3 \: \sin \varphi \right) \spc , }
\end{eqnarray*}
and thus obtain
\begin{eqnarray*}
B &=& \frac{i}{2} \left(\frac{2}{r} - \frac{T^\prime}{T} \right) G^r
	\:-\: \frac{i}{r} \left( \gamma^1 \: \cos \vartheta \:+\: \gamma^2 \:
	\sin \vartheta \: \cos \varphi \:+\: \gamma^3 \: \sin \vartheta \:
	\sin \varphi \right) \\
&=& \frac{i}{r} \:(1-A^{-\frac{1}{2}}) \: G^r \:-\: \frac{i}{2} 
    \:\frac{T^\prime}{T} \: G^r \spc .
\end{eqnarray*}
We conclude that the Dirac operator has the form
\begin{equation}
G \;=\; i G^t \frac{\partial}{\partial t} \:+\: G^r
	\left( i \frac{\partial}{\partial r} \:+\: \frac{i}{r}
	\:(1-A^{-\frac{1}{2}})\:-\: \frac{i}{2} \: \frac{T^\prime}{T} \right)
	\:+\: i G^\vartheta \frac{\partial}{\partial \vartheta}
	\:+\: i G^\varphi \frac{\partial}{\partial \varphi}
	\;\;\; . \label{8a}
\end{equation}

\section{The Dirac Equations}
\label{sec4}\setcounter{equation}{0}
In this section, we shall separate out angular momentum from the Dirac
equation (\ref{2.3a}) and reduce the problem to one on real 2-spinors.

We first introduce some formulas involving Pauli matrices. These will 
be used in this section for the separation of the angular dependence, and 
then, in the next section, for the computation of the energy-momentum tensor
needed in Einstein's equations. We introduce the following notation:
\begin{eqnarray*}
\sigma^r(\vartheta, \varphi) &=& \sigma^1 \:\cos \vartheta \:+\: \sigma^2 \:
	\sin \vartheta \: \cos \varphi \:+\: \sigma^3 \: \sin \vartheta \:
	\sin \varphi \\
\sigma^\vartheta(\vartheta, \varphi) &=& -\sigma^1 \:\sin \vartheta \:+\: \sigma^2 \:
	\cos \vartheta \: \cos \varphi \:+\: \sigma^3 \: \cos \vartheta \:
	\sin \varphi \\
\sigma^\varphi(\vartheta, \varphi) &=& \frac{1}{\sin \vartheta} \:
    (-\sigma^2 \: \sin \varphi \:+\: \sigma^3 \: \cos \varphi)
\end{eqnarray*}
These matrices are orthogonal,
\[ \Tr (\sigma^r \:\sigma^\vartheta) \;=\; \Tr (\sigma^r 
\:\sigma^\varphi) \;=\; \Tr (\sigma^\vartheta \:\sigma^\varphi) \;=\; 
0 \spc , \]
and their square is a multiple of the identity,
\[ (\sigma^r)^2 \;=\; (\sigma^\vartheta)^2 \;=\; \1 \;\;\;,\spc
(\sigma^\vartheta)^2 \;=\; \frac{\1}{\sin^2 \vartheta} \spc . \]
Furthermore,
\begin{eqnarray}
\label{d}
\sigma^\vartheta \:(\partial_\vartheta \sigma^r) &=&
\sigma^\varphi \:(\partial_\varphi \sigma^r) \;=\; \1 \\
\label{f}
\Tr \left( \sigma^\vartheta \:(\partial_\varphi \sigma^r) \right) &=&
	2 \:\sin \vartheta \: \cos \vartheta
\:(-\cos \varphi \: \sin \varphi \:+\:
	\sin \varphi \: \cos \varphi) \;=\; 0 \\
\label{g}
\Tr \left( \sigma^\varphi \:(\partial_\vartheta \sigma^r) \right) &=&
	2\:\frac{\cos \vartheta}{\sin \vartheta} \:(-\sin \varphi \: \cos \varphi
	\:+\: \cos \varphi \: \sin \varphi) \;=\; 0 \spc .
\end{eqnarray}

In analogy to the ansatz for the Dirac spinors in the hydrogen atom for
zero angular momentum (see e.g.\ \cite{10}), we write the wave functions in the 
form
\begin{equation}
	\Psi_a \;=\; e^{-i \omega t} \:
	\left( \begin{array}{c} u_1 \:e_a \\ \sigma^r \:u_2 \:e_a
	\end{array} \right) \;\;\;,\spc a=1,2 \spc ,
	\label{0}
\end{equation}
where $u_1(r)$ and $u_2(r)$ are complex-valued functions, and the
$(e_a)_{a=1,2}$ denote the standard basis $e_1=(1,0)$, 
$e_2=(0,1)$ of the two-component Pauli spinors.
This ansatz is quite useful, because the Dirac equations for $\Psi_1$ 
and $\Psi_2$ are independent of each other,
\begin{eqnarray}
G \: \Psi_a &=& \left[
    \left( \begin{array}{cc} 0 & \sigma^r \\ -\sigma^r & 0 \end{array} \right)
	\left(i \sqrt{A} \:\partial_r \:+\: \frac{i}{r} \:
	(\sqrt{A} - 1) \:-\: \frac{i}{2} \: \frac{T^\prime}{T} \:
	\sqrt{A} \right) \right. \nonumber \\
&&\left. \hspace*{3cm} \:+\: \omega \:T \: \gamma^0 \:+\:
\frac{2 i}{r} \: \left( \begin{array}{cc} 0 & \sigma^r \\ 0 & 0 \end{array}
\right) \right] \Psi_a \spc ,
\label{1}
\end{eqnarray}
where we have used (\ref{d}).
This allows us to view the Dirac equation as a two-component equation in
$u_1$, $u_2$. In order to simplify the radial dependence, we choose new 
functions $\Phi_1$ and $\Phi_2$ defined by
\begin{equation}
	\Phi_1 \;=\; r \: T^{-\frac{1}{2}} \: u_1 \;\;\;,\spc
	\Phi_2 \;=\; -i r \: T^{-\frac{1}{2}} \: u_2 \spc ,
	\label{35a}
\end{equation}
and rewrite the Dirac equation as
\begin{equation}
\left[ \left( \begin{array}{cc} 1 & 0 \\ 0 & -1 \end{array} \right)
	   \omega\: T \:-\: \left( \begin{array}{cc} 0 & 1 \\
	    1 & 0 \end{array} \right) \sqrt{A} \:\partial_r
	    \:+\: \left( \begin{array}{cc} 0 & -1 \\
	    1 & 0 \end{array} \right) \frac{1}{r} \:-\: m \right] \Phi 
	    \;=\;0 \spc .
	\label{de1}
\end{equation}
This equation is real; thus we may assume that $\Phi$ itself is real.
On these two-component spinors, the scalar product $\bra .|. \ket$ 
takes the form
\[ \bra \Phi \:|\: \Psi \ket \;=\; \int_0^\infty \overline{\Phi} 
\Psi \: A^{-\frac{1}{2}} \: dr \;\;\;,\spc \overline{\Phi} \;=\;
\Phi \sigma^3 \spc . \]
The ``Dirac operator'' in (\ref{de1}) is Hermitian with respect to 
this scalar product, and the normalization condition (\ref{nc}) for the
wave functions transforms into
\begin{equation}
\int_0^\infty |\Phi|^2 \: \frac{T}{\sqrt{A}} \; dr \;=\;
\frac{1}{4 \pi} \spc .
\label{16a}
\end{equation}
Finally, we write the Dirac equation as the ODE
\begin{equation}
	\sqrt{A} \: \Phi^\prime \;=\; \left[ \omega T
	\left( \begin{array}{cc} 0 & -1 \\ 1 & 0 \end{array} \right)
	\:+\: \frac{1}{r} \:\left( \begin{array}{cc} 1 & 0 \\
    0 & -1 \end{array} \right) \:-\: m\:
    \left( \begin{array}{cc} 0 & 1 \\
    1 & 0 \end{array} \right) \right] \Phi \spc .
	\label{3}
\end{equation}

We remark that, instead of taking the ansatz (\ref{0}) for the wave functions,
we could just as well have put the matrix $\sigma^r$ into the upper
component, i.e.
\begin{equation}
	\Psi_a \;=\; e^{-i \omega t} \:
	\left( \begin{array}{c} \sigma^r u_1 \:e_a \\ u_2 \:e_a
	\end{array} \right) \;\;\;,\spc a=1,2 \spc .
	\label{0z}
\end{equation}
This ansatz can be reduced to (\ref{0}) by changing the sign of the 
mass in the Dirac equation. To see this, we transform the wave function
according to
\[ \hat{\Psi}_a \;=\; \gamma^5 \:\Psi_a \spc . \]
Then since $\hat{\Psi}$ is of the form (\ref{0}) and satisfies the Dirac
equation
\[ 0 \;=\; \gamma^5 \:(G-m) \gamma^5 \:\hat{\Psi}_a \;=\;
-(G+m) \:\hat{\Psi}_a \spc , \]
we can again simplify to the two-component equation (\ref{de1}). We 
conclude that it also makes physical sense to look at the solutions 
of (\ref{3}) with negative $m$ (and positive $\omega$); they can be 
interpreted as solutions corresponding to the ansatz (\ref{0z}).

In Appendix \ref{asd}, we study the spinor dependence of general
static, spherically symmetric solutions of the Einstein-Dirac 
equations, and we give a more systematic justification for the 
two ansatz' (\ref{0}) and (\ref{0z}).

\section{Calculation of the Energy-Momentum Tensor}
\setcounter{equation}{0}
We derive the form of the energy-momentum tensor by considering an 
arbitrary variation $\delta g_{ij}$ of the metric tensor. The 
variation of its inverse is given by $\delta g^{ij}= -g^{ik} \:g^{jl} 
\: \delta g_{kl}$. In order to satisfy (\ref{dm}), we vary the Dirac matrices
according to
\begin{equation}
	\delta G^j \;=\; -\frac{1}{2} \:g^{jk}\:(\delta g_{kl}) \:G^l \;\;\; ,\spc
	\delta G_j \;=\; \frac{1}{2} \:(\delta g_{jk}) \:G^k \spc .
	\label{51}
\end{equation}
The energy-momentum tensor is obtained as the variation of the 
classical Dirac action (see e.g.\ \cite{4}),
\[ S \;=\; \int \overline{\Psi}(G-m) \Psi \:\sqrt{|g|} \:d^4x \spc . \]
This action is real; therefore it suffices to consider the real part of the 
integrand. Since the wave function $\Psi$ solves the Dirac equation
for the unperturbed Dirac operator $G$, we must only consider the variation
of $G$; thus
\begin{equation}
	\delta S \;=\; \int {\mbox{Re }} \overline{\Psi} \left(
		i (\delta G^j) \frac{\partial}{\partial x^j} \:+\: \delta B \right)
		 \Psi \:\sqrt{|g|} \:d^4x \spc .
	\label{52}
\end{equation}
We calculate the variation of the matrix $B$ using (\ref{51}) and
(\ref{bf}), ({\ref{gf}),
\begin{eqnarray}
{\mbox{Re }} \overline{\Psi} \:\delta B \:\Psi
& = & \frac{1}{16} \:{\mbox{Im }} \delta \left( \Tr (G^m \:\nabla_j G^n) \;
\overline{\Psi} \:G^j \:G_m G_n \Psi \right) \nonumber \\
& = & \frac{1}{16} \:\delta \left( \Tr (G^m \:\nabla_j G^n) \;
{\mbox{Im }}(\overline{\Psi} \:G^j \:G_m G_n \Psi) \right) \nonumber \\
&=& \frac{1}{16} \:\delta \left( \epsilon^{jmnp} \:\Tr (G_m \:\nabla_j G_n)
\: \overline{\Psi} \:\gamma^5 G_p\: \Psi \right) \nonumber \\
&=& \frac{1}{16} \:\delta \left( \epsilon^{jmnp} \:\Tr (G_m \:\partial_j G_n)
\: \overline{\Psi} \:\gamma^5 G_p\: \Psi \right) \nonumber \\
&\stackrel{(\ref{2.16c})}{=}& \frac{1}{16} \:\epsilon^{jmnp} \;\delta
\Tr (G_m \:\partial_j G_n) \;\overline{\Psi} \:\gamma^5 G_p\: \Psi \nonumber \\
&=&\frac{1}{16} \:\epsilon^{jmnp} \:
(\delta g_{mk}) \; \Tr (G^k \:\partial_j G_n)
\;\overline{\Psi} \:\gamma^5 G_p\: \Psi \spc . \label{42a}
\end{eqnarray}
Notice that the trace in the last equation does not necessarily vanish.
But we can use the fact that we are dealing with a spin singlet; this implies
that the expectation value of the pseudovector $\gamma^5 G_p$
is zero,
\[ \sum_{a=1}^2 \overline{\Psi_a} \:\gamma^5 G_p\: \Psi_a 
\;=\; 0 \]
(this can be checked by a short explicit calculation).
Thus we only have to consider the variation $\delta G^j$ of the Dirac 
matrices in (\ref{52}). We substitute (\ref{51}) into (\ref{52}) and obtain
for the spin singlet
\[ \delta S \;=\; \int \frac{1}{2} \sum_{a=1}^2 {\mbox{ Re}}\;
	\overline{\Psi_a} \left( i G_j \frac{\partial}{\partial x^k} 
	\right) \Psi_a \; \delta g^{jk} \; \sqrt{|g|} \:d^4 x \spc . \]
Thus the energy-momentum tensor has the form
\begin{equation}
	T_{jk} \;=\; \frac{1}{2} \sum_{a=1}^2 {\mbox{ Re}}\;
	\overline{\Psi_a} \left(i G_j \frac{\partial}{\partial x^k}
	+ i G_k \frac{\partial}{\partial x^j} \right) \Psi_a
	\spc . \label{4.3}
\end{equation}
We compute
\begin{eqnarray*}
{\mbox{Re }}
\sum_{a=1}^2 \overline{\Psi_a} \:i G^t \:\partial_t \Psi_a &=&
2 \omega \:T \:|\Psi|^2 \;=\; 2 \omega \:T^2\:
r^{-2} \: |\Phi|^2 \\
{\mbox{Re }}
\sum_{a=1}^2 \overline{\Psi_a} \:i G^t \:\partial_r \Psi_a &=&
{\mbox{Re }} 2 i T \left( r^{-1} \: \sqrt{T} \: \Phi \right) \:
\partial_r \left( r^{-1} \:\sqrt{T} \: \Phi \right) \;=\; 0 \\
{\mbox{Re }}
\sum_{a=1}^2 \overline{\Psi_a} \:i G^r\: \partial_t \Psi_a &=&
{\mbox{Re }} \omega \sum_{a=1}^2 \overline{\Psi_a} G^r \Psi_a
\;=\; {\mbox{Re }} 2 i \omega \:\frac{\sqrt{A} T}{r^2}\:
\Phi \left( \begin{array}{cc} 0 & 1  \\ -1 & 0 
\end{array} \right) \Phi \;=\; 0 \\
{\mbox{Re }} \sum_{a=1}^2 \overline{\Psi_a} \:i G^r\: \partial_r
\Psi_a &=&{\mbox{Re }}
\sum_{a=1}^2 i \sqrt{A} \: \overline{\Psi_a} \: \left( 
\begin{array}{cc} 0 & \1  \\ -\1 & 0 \end{array} \right)
\frac{\vec{x} \vec{\sigma}}{r} \: \partial_r \Psi_a \\
&=& {\mbox{Re }} 2 i \sqrt{A} \left( r^{-1}\: \sqrt{T} \: \Phi \right)
\left( \begin{array}{cc} 0 & i \\ -i & 0 \end{array} \right) 
\partial_r \left( r^{-1} \: \sqrt{T} \: \Phi \right) \\
&=& {\mbox{Re }} 2 \sqrt{A} \:T \: r^{-2} \: \Phi
\left( \begin{array}{cc} 0 & -1 \\ 1 & 0 \end{array} \right) 
\partial_r \Phi \\
&\stackrel{(\ref{3})}{=}&
-2 \omega \:T^2 \: r^{-2} \: |\Phi|^2 \:+\: 4 T
\: r^{-3} \: \Phi_1 \Phi_2 \:+\: 2m \: T \:r^{-2}\:
(\Phi_1^2 - \Phi_2^2) \\
{\mbox{Re }} \sum_{a=1}^2 \overline{\Psi_a} \:i G^\vartheta\:
\partial_\vartheta \Psi_a &=&{\mbox{Re }}
r^{-1} \:\sum_{a=1}^2 \overline{\Psi_a} \:
\left( \begin{array}{cc} 0 & i \1 \\ -i \1 & 0 \end{array} \right)
\sigma^\vartheta\: \partial_\vartheta \Psi_a \\
&\stackrel{(\ref{d})}{=}& -2 r^{-3} \:T \: \Phi_1 \Phi_2 \\
{\mbox{Re }} \sum_{a=1}^2 \overline{\Psi_a}
\:i G^\varphi\: \partial_\varphi \Psi_a &=&
{\mbox{Re }} r^{-1} \:\sum_{a=1}^2 \overline{\Psi_a} \:
\left( \begin{array}{cc} 0 & i \1 \\ -i \1 & 0 \end{array} \right)
\sigma^\varphi\: \partial_\varphi \Psi_a \\
&\stackrel{(\ref{d})}{=}& -2 r^{-3} \: T \: \Phi_1 \Phi_2 \spc .
\end{eqnarray*}
All other combinations of the indices vanish because of 
(\ref{f}), (\ref{g}), and the orthogonality of $\sigma^r$, 
$\sigma^\vartheta$, $\sigma^\varphi$. We conclude that
\begin{eqnarray}
T^i_j & = & r^{-2} \;{\mbox{diag }} \left(
	2 \omega \:T^2\: |\Phi|^2, \;
	-2 \omega \:T^2\: |\Phi|^2 + 4 T \: r^{-1} \:
	\Phi_1 \Phi_2 \:+\: 2 m \:T \:(\Phi_1^2 - \Phi_2^2),
	\right. \nonumber \\
&&\hspace*{3cm} \left. -2 T \:r^{-1} \:\Phi_1 \Phi_2,\;
	-2 T \:r^{-1} \:\Phi_1 \Phi_2 \right) \label{c}
	\spc .
\end{eqnarray}
As a check, we calculate the trace of $T^i_j$ directly from 
(\ref{4.3}),
\begin{eqnarray*}
	T^j_j & = & \sum_{a=1}^2 {\mbox{Re }} \left( 
	\overline{\Psi_a} (i G^j \partial_j) \Psi_a \right)
	\;=\; \sum_{a=1}^2 {\mbox{Re }} \left( 
	\overline{\Psi_a} (i G^j \partial_j + B) \Psi_a \right)  \\
	 & = & m \sum_{a=1}^2 \overline{\Psi_a} \Psi_a \;=\;
	 2 m \:T\:r^{-2} \: (\Phi_1^2 - \Phi_2^2) \spc ,
\end{eqnarray*}
and we see that it agrees with taking the trace of (\ref{c}).

\section{The Field Equations}
\setcounter{equation}{0}
The Einstein Tensor $G^i_j = R^i_j - \frac{1}{2} R \:\delta^i_j$ has the form
(see e.g.\ \cite{1})
\begin{eqnarray*}
G^0_0 &=&-\frac{1}{{r^2}}+\frac{A}{{r^2}}+\frac{{A^{\prime }}}{r} \\
G^1_1 &=&-\frac{1}{{r^2}}+\frac{A}{{r^2}}-\frac{2 A {T^{\prime }}}{r T}\\
G^2_2 &=& G^3_3 \;=\; \frac{{A^{\prime }}}{2 r}-\frac{A {T^{\prime }}}{r T}
-\frac{{A^{\prime }} {T^{\prime }}}{2 T}+\frac{2 A
{{{T^{\prime }}}^2}}{{{T}^2}}-\frac{A {T^{\prime \prime}}}{T} \spc ,
\end{eqnarray*}
with all other components vanishing.
Thus, using (\ref{c}), Einstein's field equations $G^i_j=-8 \pi \:T^i_j$ become
\begin{eqnarray}
\label{x}
-16 \pi \omega \: T^2 \:|\Phi|^2 &=& r\: A^\prime
- (1-A) \\
\label{y}
\lefteqn{ \hspace*{-3cm} -16 \pi \omega \:T^2\: |\Phi|^2
+ 32 \pi r^{-1} \:T \:\Phi_1 \Phi_2 \:+\: 16 \pi m \:T \: (\Phi_1^2 - \Phi_2^2)
\;=\; 2r A \:\frac{T^\prime}{T} \:+\: (1-A) } \\
\label{z}
-16 \pi T \:r^{-1} \:\Phi_1 \Phi_2
&=& A \left[ r^2 \:\frac{T^{\prime \prime}}{T} + r^2
\: \frac{A^\prime T^\prime}{2 \:A T}
- 2 r^2 \left( \frac{T^\prime}{T} \right)^2
- r \: \frac{A^\prime}{2 A} \:+\: r \: \frac{T^\prime}{T} \right] 
\;\;\; . \spc
\end{eqnarray}

Using the notation $\alpha=\Phi_1$, $\beta=\Phi_2$, the equations 
(\ref{3}), (\ref{x}), and (\ref{y}) can be written as
\begin{eqnarray}
	\sqrt{A} \:\alpha^\prime & = & \frac{1}{r} \:\alpha \:-\: (\omega T + 
	m) \:\beta \label{dirac1} \\
	\sqrt{A} \:\beta^\prime	 & = & (\omega T - m) \:\alpha \:-\:
	\frac{1}{r} \:\beta \label{dirac2} \\
	r \:A^\prime & = & 1-A \:-\: 16 \pi \omega T^2 \:(\alpha^2 + \beta^2)
	\label{Aeq} \\
	2 r A \:\frac{T^\prime}{T} & = & A-1 \:-\: 16 \pi \omega T^2 
	\:(\alpha^2+\beta^2) \:+\: 32 \pi \:\frac{1}{r} \:T\: \alpha \beta \:+\:
	16 \pi \:m T \:(\alpha^2-\beta^2) \;\;\; . \spc
	\label{Teq}
\end{eqnarray}
By direct computation (or, e.g.\ by using Mathematica), we see that 
equation (\ref{z}) is implied by this set of equations.
It is sometimes useful to rewrite the Dirac equations 
(\ref{dirac1}) and (\ref{dirac2}) in matrix notation as
\[ \sqrt{A} \:\Phi^\prime \;=\; \left( \begin{array}{cc}
1/r & -\omega T - m \\ \omega T - m & -1/r 
\end{array} \right) \Phi \spc . \]
The normalization condition (\ref{nc}) takes the form
\begin{equation}
\int_0^\infty |\Phi|^2 \:\frac{T}{\sqrt{A}} \; dr \;=\; \frac{1}{4 \pi} \spc .
	\label{norm}
\end{equation}

If we assume regularity of the solution at $r=0$, we obtain the 
following Taylor series expansions
\begin{eqnarray}
\alpha(r) &=& \alpha_1 \:r \:+\: {\cal{O}}(r^3) \label{ic1} \\
\beta(r) &=& \frac{1}{3} \:(\omega T_0 - m) \:\alpha_1 \:r^2
\:+\: {\cal{O}}(r^3) \label{ic2} \\
A(r) &=& 1 \:-\: \frac{2}{3} \:\omega T_0^2 \alpha_1^2 \:r^2 \:+\: 
{\cal{O}}(r^3) \label{ic3} \\
T(r) &=& T_0 \:-\:\frac{m}{6} \:(4 \omega T_0 - 3m) \:T_0^2 
\:\alpha_1^2 \:r^2 \:+\:{\cal{O}}(r^3) \spc , \label{ic4} 
\end{eqnarray}
where $\alpha_1$, $T_0$, $\omega$ and $m$ are (real) parameters.
We also require that our solutions have finite (ADM) mass; namely
\begin{equation}
	\rho \;:=\; \lim_{r \rightarrow \infty} \:\frac{r}{2} \:(1-A(r))
	\;<\; \infty \spc .
	\label{5.14}
\end{equation}
This implies that
\begin{equation}
	\lim_{r \rightarrow \infty} A(r) \;=\; 1 \spc .
	\label{asy1}
\end{equation}
Finally, in order that the metric be asymptotically Minkowskian, we 
require that
\begin{equation}
\lim_{r \rightarrow \infty} T(r) \;=\; 1 \spc . \label{asy2}
\end{equation}

\section{Scaling of the Equations}
\label{sec8} \setcounter{equation}{0}
For the numerical computations, it is difficult to take into account condition 
(\ref{asy2}) and the integral condition (\ref{norm}).
Therefore we find it convenient to temporarily replace these 
constraints by the weaker conditions
\begin{equation}
	\lim_{r \rightarrow \infty} T(r) \;<\; \infty \spc {\mbox{and}} \spc
	\int_0^\infty |\Phi|^2 \:\frac{T}{\sqrt{A}} \; dr \;<\; \infty
	\label{s1weak}
\end{equation}
and instead set
\begin{equation}
	T_0 \;=\; 1 \spc {\mbox{and}} \spc m \;=\; \pm 1 \spc .
	\label{s1}
\end{equation}
This is justified by the following scaling argument which shows that
there is a one-to-one correspondence between solutions satisfying
(\ref{s1weak}),(\ref{s1}) and solutions satisfying (\ref{asy2}) and (\ref{norm}).

Thus, suppose we have a solution $(\alpha, \beta, T, A)$ of 
(\ref{dirac1})-(\ref{Teq}), (\ref{5.14}) satisfying
(\ref{s1weak}),(\ref{s1}). Then, with the parameters $\lambda$ and $\tau$ 
defined by
\begin{eqnarray*}
\lambda &=& \left( 4 \pi \int_0^\infty (\alpha^2+\beta^2) \: \frac{T}{\sqrt{A}} 
\:dr \right)^{\frac{1}{2}} \\
\tau &=& \lim_{r \rightarrow \infty} T(r) \spc ,
\end{eqnarray*}
we introduce the new functions
\begin{eqnarray*}
\tilde{\alpha}(r) & = & \sqrt{\frac{\tau}{\lambda}}
	\:\alpha(\lambda r) \\
\tilde{\beta}(r) & = & \sqrt{\frac{\tau}{\lambda}}
	\:\beta(\lambda r) \\
\tilde{A}(r) &=& A(\lambda r) \\
\tilde{T}(r) &=& \tau^{-1} \:T(\lambda r) \spc .
\end{eqnarray*}
A direct calculation shows that these functions satisfy the equations
(\ref{dirac1})-(\ref{Teq}) with $m, \omega$ replaced by
\[ \tilde{m} \;=\; \lambda m \;\;\;,\spc \tilde{w} \;=\; 
\lambda \:\omega \tau \spc . \]
Moreover, it is easy to check that
\begin{eqnarray*}
\int_0^\infty (\tilde{\alpha}^2+\tilde{\beta}^2) 
\:\frac{\tilde{T}}{\sqrt{\tilde{A}}} \:dr &=& \frac{1}{4 \pi} \spc , \\
\lim_{r \rightarrow \infty} \tilde{T}(r) &=& 1
\end{eqnarray*}
and $\lim_{r \rightarrow \infty} \frac{r}{2} \:(1-\tilde{A}(r)) < \infty$.
We note that, as long as $\lambda, \tau$ are positive and non-zero, 
the transformation from the un-tilde variables to the tilde variables 
is one-to-one.

Finally, we point out that for the physical interpretation, the 
conditions (\ref{norm}) and (\ref{asy2}) must hold and hence {\em{only the 
scaled solutions can have physical significance}}.

\section{Existence and Properties of the Solutions}
\setcounter{equation}{0} \label{sec9}
Using the local Taylor expansion about $r=0$, 
(\ref{ic1})-(\ref{ic4}), and setting $T(0)=1$ and $m=\pm1$, we construct initial
data at $r=10^{-5}$ and then use the standard Mathematica ODE solver.
We shoot for numerical solutions by fixing $\alpha_1$ and varying 
$\omega$. Using accuracy $10^{-8}$, we found that solutions varied 
continuously with $\alpha_1$, (and $\omega$), indicating that this 
procedure is well-behaved. The solutions we find satisfy
\begin{eqnarray*}
\lim_{r \rightarrow \infty} r^2 \:(\alpha(r)^2+\beta(r)^2) &=& 0
\;\;\;\;\:,\spc \lim_{r \rightarrow \infty} A(r) \;=\; 1 \\
\lim_{r \rightarrow \infty} \frac{r}{2} \:(1-A(r)) &<& \infty \;\;\;,\spc
\lim_{r \rightarrow \infty} T(r) \;=\; \tau \;>\; 0
 \spc ,
\end{eqnarray*}
and, in addition, $T$ and $A$ stay positive for all $r$. In order to 
fulfill the important other two conditions (\ref{norm}),(\ref{asy2}),
we merely scale our variables in the manner described in section 
\ref{sec8}.

For any given $\alpha_1>0$, we found a unique ground state corresponding
to the parameter value $\omega_0$, together with a countable number
of distinct excited states corresponding to parameter values $\omega_n$,
$n=1,2,\ldots$, where 
$\omega_0<\omega_1<\ldots<\omega_{\mbox{\scriptsize{max}}}(\alpha_1)$.
For $\alpha_1=0.02$ and $m=1$, the {\em{scaled}} solutions for the 
ground state and for the first and second excited states are depicted in
Figures \ref{ground}, \ref{first} and \ref{second}\footnote{We point 
out that we also plot the {\em{scaled}} variables in all the other 
figures.}.
These solutions have the following general characteristics:
The graphs of the functions $T(r)$ have the same qualitative form for 
all values $\omega_n$. The functions $A(r)$, however, 
have their graphs changing with $\omega_n$; indeed, for the $n$th excited
state, the graph of $A$ has precisely $n+1$ minima.
The tangent vector to the curve $(\alpha(r), \beta(r))$ for large $r$ lies
alternately in either the first or third quadrants.

For larger values of $\alpha_1$ (and $m=1$), the solutions have a 
similar form, but the $\alpha-\beta$ graphs of the ground state have
self-intersections and are thus of a different shape. This is 
illustrated for three different values of $\alpha_1$ in Figure 
\ref{loop}.
We found that for the ground state,
\[ \lim_{\alpha_1 \rightarrow \infty} T_0(\alpha_1) \;=\; \infty
\spc {\mbox{and}} \spc \lim_{\alpha_1 \rightarrow 0} T_0(\alpha_1)
\;=\; 1 \spc . \]
Moreover, in every case, $T$ is a monotone decreasing 
function of $r$.

We also obtained solutions for $m=-1$ by using similar methods. The 
results are qualitatively the same with the exception
that, in this case, the tangent vector to the $(\alpha, \beta)$ curve
for large $r$ lies alternatively in quadrants two and four,
see Figures \ref{nfirst} and \ref{nsecond}.

The mass and the energy of the solutions we found have some interesting and
surprising features, which we shall now describe.
We consider the fixed $n^{\mbox{\tiny{th}}}$ excited state and, by 
varying $\alpha_1$ and adjusting $\omega$ (for fixed $m=\pm1$), we obtain a 
one-parameter family of solutions (parametrized e.g.\ by $\alpha_1$).
After scaling, we find that solutions can only exist for a bounded 
range of $m$. For every value of $m$ in this range, we obtain an at 
most countable number of solutions, which can be characterized by two physical 
parameters: the energy $\omega$ of the fermions, 
and the (ADM) mass (\ref{5.14}). We find that we always have 
$\omega < |m|$, which means that the fermions are in a bound 
state. If we plot the binding energy $\omega - |m|$ vs.\ the mass 
$m$ (i.e., the mass-energy spectrum), we find that this curve is a spiral
which tends to a limiting configuration. This is shown in Figure 
\ref{bind} for the ground state (G), the first and second excited 
states (F, S), and the lowest and next excited states for negative 
mass (L, N).
The spirals seem to have a self-similarity; this is illustrated in
Figure \ref{zoom} (we stopped the computation when the limitations of our 
numerics were reached).
The (ADM) mass $\rho$ can be viewed as the total energy of both the 
gravitational field and the fermions. Thus the quantity $\rho - 2 
\omega$ gives the energy of the gravitational field. As 
is shown in Figure \ref{adm}, it is always positive and, if plotted 
versus $m$, gives curves which again look like spirals.
Furthermore, one can ask how much energy is gained (or needed) to form 
the singlet state. For this, we must compare the (ADM) mass with 
the total rest mass $2 |m|$; i.e., the energy of two fermions at rest 
which are so much delocalized in space that their gravitational 
interaction becomes zero. This is plotted in Figure \ref{tm}.
For small $m$, the function $\rho - 2 |m|$ is negative, which shows that one
gains energy by forming the singlet state. (This gives a first hint
that these states might be stable, because energy is needed to break
up the binding.) For large values of $m$, however, (more precisely, shortly 
after the ``turning point'' of $m$; see Figure \ref{tm}), $\rho - 2 |m|$
becomes positive. 
This indicates that the solutions should no longer be stable. This is 
indeed true and will be shown in the next section.

We note that our scaling technique is essential for obtaining the mass
spectrum - the unscaled variables do not have ``spirals''.

\section{Stability of the Solutions}
\label{sec10}
\setcounter{equation}{0}
In this section, we shall consider the stability of our solutions 
under spherically symmetric perturbations.
To this end, we consider the spherically symmetric, time-dependent
Lorentzian metric of the form
\[ ds^2 \;=\; T^{-2}(r,t) \:dt^2 \:-\: A^{-1}(r,t) \:dr^2 \:-\: 
r^2 \:(d\vartheta^2 \:+\: \sin^2 \vartheta \: d\varphi^2) \spc . \]
The time-dependent Dirac operator can again be calculated using
(\ref{bf1}). One gets the following generalization of (\ref{8a})
(the dot denotes t-derivatives):
\[ G \;=\; G^t \left( i \frac{\partial}{\partial t} \:-\:
\frac{i}{4} \: \frac{\dot{A}}{A} \right)
\:+\: G^r \left( i \frac{\partial}{\partial r} \:+\: \frac{i}{r}
	\:(1-A^{-\frac{1}{2}})\:-\: \frac{i}{2} \: \frac{T^\prime}{T} \right)
	\:+\: i G^\vartheta \partial_\vartheta + i G^\varphi \partial_\varphi
	\spc . \]
As in Section \ref{sec4}, we separate out the angular momentum by
setting (cf.\ (\ref{0}))
\[ \Psi \;=\; \frac{\sqrt{T}}{r} \:\left( \begin{array}{c}
z_1(r,t) \:e_a \\ i \:\sigma^r \: z_2(r,t) \:e_a
\end{array} \right) \;\;\;,\spc a=1,2 \spc . \]
This gives the two-component, time-dependent Dirac equation
\begin{eqnarray*}
\lefteqn{ \left[ \left( \begin{array}{cc} 1 & 0 \\ 0 & -1 \end{array} \right)
	   \left(i T \partial_t \:-\: \frac{i}{4} \:T\: \frac{\dot{A}}{A}
	   \:+\: \frac{i}{2} \:
	  \dot{T} \right) \right. } \\
&&\hspace*{2cm} \left.
	   \:-\: \left( \begin{array}{cc} 0 & 1 \\
	    1 & 0 \end{array} \right) \sqrt{A} \:\partial_r
	    \:+\: \left( \begin{array}{cc} 0 & -1 \\
	    1 & 0 \end{array} \right) \frac{1}{r} \:-\: m \right]
	    \left( \begin{array}{c} z_1 \\ z_2 \end{array} \right)
	    \;=\; 0 \spc .
\end{eqnarray*}
Observe that, in contrast to (\ref{de1}), this equation is complex, and 
thus we cannot assume here that the spinors are real.
Calculation of the energy-momentum tensor and the Einstein tensor 
gives the equations
\begin{eqnarray*}
-\frac{1}{{r^2}}+\frac{A}{{r^2}}+\frac{A^\prime}{r} &=& -8 \pi \;
\frac{2 i T^2}{r^2} {\mbox{ Re}} \left( \overline{z_1}
\:{{\partial }_t}{z_1} +\overline{z_2}\:{{\partial }_t}{z_2} \right) \\
\frac{T^2 \dot{A}}{r A} &=& -8 \pi {\mbox{ Re}} \left(
\frac{i T^2}{r^2} \:(\overline{z_1}\: \partial_r z_1
+ \overline{z_2} \:\partial_r z_2) \:+\: \frac{T^3 A^{-\frac{1}{2}}}{{r^2}} \:
(\overline{z_1}\: \partial_t z_2 -\overline{z_2} \:\partial_t z_1) \right) \\
-\frac{1}{{r^2}}+\frac{A}{{r^2}}-\frac{2 A T^\prime}{r T} &=& 8 \pi \;
\frac{2 T A^\frac{1}{2}}{{r^2}} {\mbox{ Re}}
\left( \overline{z_1}\: \partial_r z_2 - \overline{z_2} \: \partial_r 
z_1 \right) \\
\lefteqn{ \hspace*{-3cm} \frac{A^\prime}{2 r}-\frac{A T^\prime}{r T}
-\frac{A^\prime T^\prime}{2}
+\frac{2 A T^\prime}{T^2}-\frac{A T^{\prime \prime}}{T}
-\frac{3 T^2 \dot{A}^2}{4 A^2}+\frac{T \dot{A} \dot{T}}{2 A}
+\frac{T^2 \ddot{A}}{2 A} \;=\;
8 \pi \;\frac{2}{{r^3}}\:T\: {\mbox{ Re}} (\overline{z_1} z_2) \spc . }
\end{eqnarray*}

Suppose that a solution $(\alpha(r), \beta(r), A(r), T(r))$ of the equations
(\ref{dirac1})-(\ref{Teq}) and (\ref{norm}), (\ref{asy1}), (\ref{asy2})
is given for some values of the parameters $m$, $\omega$.
Before making an ansatz for a time-dependent perturbation of this 
solution, it is convenient to introduce new spinor variables $\alpha(r,t)$, 
$\beta(r,t)$ by setting
\begin{equation}
	z_1(r,t) \;=\; e^{-i \omega t} \: \alpha(r,t) \;\;\;,\spc
	z_2(r,t) \;=\; e^{-i \omega t} \: \beta(r,t) \spc .
	\label{F2}
\end{equation}
This has the advantage that we get the ansatz (\ref{0}) simply by 
assuming that $\alpha$ and $\beta$ are time independent.
We consider spherically symmetric, time-dependent perturbations of the form
\begin{eqnarray}
	\alpha(r,t) & = &
	\alpha(r) \:+\: \varepsilon (a_1(r,t)+i \:a_2(r,t)) \label{F} \\
	\beta(r,t) & = &
	\beta(r) \:+\: \varepsilon (b_1(r,t)+i \:b_2(r,t)) \\
	A(r,t) & = & A(r) \:+\: \varepsilon A_1(r,t) \\
	T(r,t) & = & T(r) \:+\: \varepsilon T_1(r,t) \spc ,
	\label{G}
\end{eqnarray}
where we look at the real and imaginary parts separately ($a_j$ and $b_j$ 
are real functions).
Substituting into the Einstein-Dirac equations and only 
considering the first order terms in $\varepsilon$ gives (using 
Mathematica) a system of linear differential equations.
If these equations admitted solutions with an exponentially growing time 
dependence, the original solution would be unstable. 
Therefore we separate out the time dependence by writing
\begin{eqnarray}
a_j(r,t) &=& e^{\kappa t} \:a_j(r) \;\;,\spc
b_j(r,t) \;=\; e^{\kappa t} \:b_j(r) \label{D} \;\;\;,\;\; j=1,2, \\
A_1(r,t) &=& e^{\kappa t} \:A_1(r) \;\;,\spc\!\!
T_1(r,t) \;=\; e^{\kappa t} \:T_1(r) \spc . \label{E}
\end{eqnarray}
This gives the following system of five ordinary differential 
equations
\begin{eqnarray}
{\sqrt{A}} \: a_1^\prime
&=&\frac{a_1}{r}-(m+\omega T) \:b_1+\kappa  T\:b_2-\frac{A_1}{2A}
\left(\frac{\alpha}{r}-(m+\omega T) \:\beta \right) -\omega \:T_1\: \beta
\label{H} \\
\sqrt{A} \:a_2^\prime
&=& \frac{a_2}{r}-(m + \omega T) \: b_2-\kappa T\:b_1+
\kappa \:\frac{A_1}{4 A}\:T \:\beta-\kappa\:\frac{T_1}{2} \:\beta \\
{\sqrt{A}} \:b_1^\prime
&=&-(m-\omega T)\: a_1-\frac{b_1}{r}-\kappa  T\:a_2
-\frac{A_1}{2A} \left(-(m-\omega T) \:\alpha - \frac{\beta}{r} \right)
+\omega  \:T_1\: \alpha \\
{\sqrt{A}} \: b_2^\prime 
&=&-(m-\omega T)\: a_2-\frac{b_2}{r}+\kappa  T\:a_1-\kappa \:
\frac{A_1}{4 A}\:T \alpha+\kappa \: \frac{T_1}{2} \:\alpha \\
2 r A T_1^\prime
&=&\frac{A_1 T}{A}-T_1+ A T_1
\:+\:\frac{32 \pi {{T}^2} }{r} \:(a_1 \beta + b_1 \alpha) \nonumber \\
&&+16 \pi T^2 \:\alpha \:(2 m a_1 -2 \omega T a_1+\kappa T a_2)
\:-\:16 \pi T^2 \beta \:(2 m b_1 + 2 \omega T b_1 - \kappa T b_2) 
\nonumber \\
&&-16 \pi \:T_1 \left( 3 \omega T^2 \:(\alpha^2+\beta^2) \:-\: 
\frac{4}{r} \:T\:\alpha \beta \:-\: 2m T \:(\alpha^2 - \beta^2) 
\right) \nonumber \\
&&+16 \pi \:\frac{A_1 T}{A} \left(
w T^2 \:(\alpha^2+\beta^2) - \frac{2}{r} \:T\:\alpha \beta - mT 
\:(\alpha^2-\beta^2) \right) \label{K}
\end{eqnarray}
together with the algebraic equation
\begin{equation}
A_1 \;=\; 16 \pi\:\frac{\sqrt{A} T}{\kappa r}\:\left( -(\kappa b_1+2 
\omega b_2) \:\alpha \:+\: (\kappa a_1+2 \omega a_2) \:\beta \right) \spc .
\label{Ka}
\end{equation}
The consistency of these equations (i.e., that the equation
$G^2_2=-8 \pi T^2_2$ is identically satisfied) was again checked with
Mathematica. We want to show that there are no solutions for $\kappa > 0$;
this implies stability.

The above equations come with initial conditions at $r=0$ and 
additional constraints, which we will now describe. A Taylor expansion 
about $r=0$ gives, similar to (\ref{ic1})-(\ref{ic4}), the following 
expansions near $r=0$
\begin{eqnarray}
a_1(r) &=& a_{10} \:r\:+\: {\cal{O}}(r^2) \;\;\;\:,\spc
a_2(r) \;=\; a_{20} \:r\:+\: {\cal{O}}(r^2) \label{L} \\
b_1(r) &=& {\cal{O}}(r^2) \spc\;\;\;\;\;\;\:,\spc b_2(r) \;=\; {\cal{O}}(r^2)
\label{M} \\
A_1(r) &=& {\cal{O}}(r^2) \spc\;\;\;\;\;\;\:,\spc T_1(r) \;=\; T_{10} \:+\: 
{\cal{O}}(r^2) \spc .
\end{eqnarray}
We have three parameters $a_{10}$, $a_{20}$ and $T_{10}$ to 
characterize the solutions.
Since the metric must be asymptotically Minkowskian, we demand 
moreover that
\begin{eqnarray}
	\lim_{r \rightarrow \infty} A_1(r) &= & 0
	\label{A}  \\
	\lim_{r \rightarrow \infty} T_1(r) & = & 0 \spc .
	\label{B}
\end{eqnarray}
Furthermore, the wave functions must be normalized, which means that
(cf.\ (\ref{16a})),
\begin{equation}
	\int_0^\infty (\alpha^2(r,t) +\beta^2(r,t)) 
	\:\frac{T(r,t)}{\sqrt{A(r,t)}} \:dr \;=\; \frac{1}{4 \pi} \spc ,
	\label{C}
\end{equation}
for all $t$. This time-dependent normalization condition appears to 
make the analysis very complicated. It turns out, however, that we do 
not have to consider it at all, because, for the perturbation 
(\ref{D}), (\ref{E}), it is automatically satisfied. 

To see this, note that, as a consequence of the current conservation
(\ref{2.5a}), the normalization integral (\ref{C}) is actually
time-independent. But, in the limit $t \rightarrow -\infty$, the
time-dependent solution (\ref{F})-(\ref{G}) of the Einstein-Dirac equations
goes over into the static, unperturbed solution $(\alpha, \beta, A, 
T)$, and thus (\ref{C}) holds in this limit. It follows that (\ref{C})
holds for all $t$. In other words, the linear contribution in $\varepsilon$ to 
(\ref{C}) vanishes as a consequence of the linearized Einstein-Dirac 
equations (\ref{H})-(\ref{K}), (\ref{Ka}). We must only make sure that 
the integral (\ref{C}) is finite for all $t$.

These conditions can be simplified. To see this, we first consider 
the infinitesimal time reparametrization
\[ t \;\rightarrow\; t \:-\: \varepsilon \:\frac{T_1(r=0)}{\kappa \:T(r=0)} \: 
e^{\kappa t} \spc . \]
This transformation does not destroy the ansatz (\ref{F})-(\ref{E}); it 
only changes the functions $a_2, b_2$ and $T_1$ according to
\begin{eqnarray}
	T_1(r) & \rightarrow & T_1(r) \:-\: T_1(0) \:\frac{T(r)}{T(0)} 
	\nonumber \\
	a_2(r) & \rightarrow & a_2(r) \:-\: 
	\frac{\omega}{\kappa}\:\frac{T_1(0)}{T(0)}\:\alpha(r)
	\label{818a}  \\
	b_2(r) & \rightarrow & b_2(r) \:-\: 
	\frac{\omega}{\kappa}\:\frac{T_1(0)}{T(0)}\:\beta(r)
	\label{818b} \spc .
\end{eqnarray}
Thus we can arrange that $T_1$ vanishes at the origin,
\[ T_1(r) \;=\; {\cal{O}}(r) \spc , \]
provided that (\ref{B}) is replaced by the weaker condition
\[ \lim_{r \rightarrow \infty} T_1(r) \;=\; \mu \spc 
{\mbox{for some $\mu$, $0 < \mu < \infty$}} \spc . \]
This makes the numerics easier, because we now have only two free 
parameters $a_{10}$, $a_{20}$ at the origin to characterize the 
solution. Furthermore, using the linearity of the equations, we can multiply
every solution by a (non-zero) arbitrary real number. This allows us to fix
one of the parameters (e.g.\ by setting $a_{20}=1$), and thus we end up 
with only one free parameter.

Our strategy is to show that, for any $\kappa >0$, there are no 
solutions for which the normalization integral (\ref{C}) is finite;
this will imply stability.
In order to explain the technique and the difficulty for the 
numerics, we consider Figure \ref{stab1}, where typical plots of 
$(a_1, b_1)$, $(a_2, b_2)$ for a small value of $\kappa$, and the 
ground state solution of Figure \ref{ground}, are shown.
According to (\ref{L}),(\ref{M}), both the $(a_1,b_1)$ and the 
$(a_2, b_2)$ curves start at the origin. We want to show that at least 
one of these curves stays bounded away from the origin as $r \rightarrow \infty$,
no matter how we choose $\kappa$ and the initial values $a_{10}$, $a_{20}$.
This will imply that the integral (\ref{C}) is unbounded.
Figure \ref{stab1} is interesting because it almost looks as if 
this happened: the $(a_2, b_2)$ curve looks like the 
$\alpha-\beta$-plot of the ground state, whereas $(a_1, b_1)$ is 
almost like the $\alpha-\beta$-plot of the first excited state.
The $(a_1, b_1)$ curve is not quite closed, however, and it is 
difficult to decide whether this is just an artifact of the numerics 
or whether it actually means that there is no normalizable solution.
The numerics are especially delicate because, for small values 
of $\kappa$, the values of $(a_2, b_2)$ are much larger than $(a_1, b_1)$ 
(notice the different scales on the plots of Figure \ref{stab1}).
In order to improve the accuracy of the numerics, it is useful 
to note that, numerically, $a_2$ and $b_2$ are, to a very good approximation, 
multiples of the unperturbed wave functions $\alpha, \beta$, i.e.\
$a_2 \approx \mu \alpha$, $b_2 \approx \mu \beta$ for some real 
constant $\mu$. We can eliminate this dominant contribution to $a_2, 
b_2$ by introducing new variables
\[ \hat{a}_2 \;=\; a_2 - \mu\: \alpha \;\;\;,\spc
\hat{b}_2 \;=\; b_2 \:-\: \mu\: \beta \]
and rewriting our ODEs in the functions $(a_1, b_1, \hat{a}_2,
\hat{b}_2, A_1, T_1)$; this gives a system of five inhomogeneous, linear 
equations. From these, we obtain the plots in Figure \ref{stab2}.
The $(a_1, b_1)$-curve is similar to that of Figure \ref{stab1}; the 
$(\hat{a}_2, \hat{b}_2)$-plot, however, gives a much more detailed 
view of the imaginary part of $\alpha(r,t), \beta(r,t)$ (notice again 
the different scales on the plot).

Next we describe our method to determine the initial data at $r=0$ 
for the solutions (i.e. the parameters $a_{10}$, $a_{20}$, for the original
equations (\ref{H})-(\ref{K})). It turns out that the integral 
(\ref{C}) is only finite for all $t$ if both $(a_1, b_1)$ and $(a_2, 
b_2)$ tend to zero for large $r$; indeed, from the numerics we see 
that $(a_1, b_1)$ and $(a_2, b_2)$ cannot have infinite oscillations.
It is an efficient technique to fix the
initial values by trying to satisfy the first part of these conditions
\begin{equation}
	\lim_{r \rightarrow \infty} (a_1(r),\: b_1(r)) \;=\; (0,0) \spc .
	\label{Z}
\end{equation}
This can be done by varying the initial values in such a way that
\[ (a_1(R)^2 + a_2(R)^2) \;\rightarrow\; {\mbox{min}} \spc , \]
where $R$ is the value of $r$ where we stop the numerics ($R$ must be 
chosen sufficiently large).
Using the linearity of our ODE's, this minimizing condition leads to 
simple algebraic relations between the initial data and the values 
of $a_1(R)$, $b_1(R)$ for a fundamental set of solutions, from which
the initial values can be computed numerically.
According to Figures \ref{stab1} and \ref{stab2}, we already know 
qualitatively how the resulting $(a_1, b_1)$-plot is supposed to look 
like. This is helpful for checking the numerical results and for 
finding the best values for $R$.
(More precisely, the best value for $R$ is, roughly speaking, the value where
the $(a_1, b_1)$-plot starts to look like a closed curve. If $R$ is 
chosen much larger than this, the numerical inaccuracies pile
up in a such way that our method of computing the initial data from $a_1(R)$
and $b_1(R)$ is no longer reliable.)

This procedure can be carried out for different values of $\kappa$, 
and gives the plots of Figures \ref{sg1} to \ref{sg6}.
For very small values of $\kappa$, the plots look like those in Figure 
\ref{sg1}, and one sees that $(\hat{a}_2(r), \hat{b}_2(r))$ does not 
tend to zero for large $r$. The shape of the $(\hat{a}_2, \hat{b}_2)$-plot 
does not change over many magnitudes in $\kappa$ (see Figures 
\ref{stab2}, \ref{sg2}, \ref{sg3}), showing too that our numerics 
are well-behaved. For $\kappa \approx 0.02$, the form of the plots 
changes drastically (see Figures \ref{sg4}, \ref{sg5}, \ref{sg6}).
One sees that $(\hat{a}_2, \hat{b}_2)$ still do not go to the origin and 
that it becomes impossible to satisfy even condition (\ref{Z}). If 
$\kappa$ is further increased, both $(a_1, b_1)$ and $(\hat{a}_2, 
\hat{b}_2)$ go to infinity for large $r$, no matter how the initial 
data is chosen. From this, we conclude that the ground state solution 
of Figure \ref{ground} is linearly stable.

Our method also applies to the excited states. For the first excited 
state and the lowest negative-mass state (i.e., for the solutions of 
the Figures \ref{first} and \ref{nfirst}), the solutions of the 
linearized equations for small $\kappa$ are plotted in Figures 
\ref{stab3} and \ref{stab4}. The $(a_2, b_2)$-curves are again of 
similar shape as the corresponding $\alpha-\beta$-plot; the $(a_1, 
b_1)$-curve resembles the $\alpha-\beta$-plot for the next higher
excited state (i.e., in Figure \ref{stab3} for the second excited 
state, and in Figure \ref{stab4} for the first excited negative-mass 
state). It is again useful to introduce the variables $\hat{a}_2$, 
$\hat{b}_2$. A numerical analysis of the equations for different 
values of $\kappa$ shows that these solutions are also stable.

It might seem a bit surprising that even the excited states are linearly 
stable. Actually, this can already be understood qualitatively from 
Figure \ref{stab3} and the following heuristic argument.
Thus, if the solution (for small coupling) were unstable, then the solution
$(a_j, b_j, A_1, T_1)$ of the linearized equations (\ref{H})-(\ref{Ka})
would, to first order, describe the decay of the bound state.
Therefore the $(a_j, b_j)$-plots give 
us information into which state the wave functions tend to decay.
The $(a_2, b_2)$-plot is not interesting in this respect, because it 
looks like the $\alpha-\beta$-plot and only yields information about a
time-dependent phase transformation of the wave functions. The $(a_1, 
b_1)$-plot, however, is helpful.
According to Figure \ref{stab3}, the wave function tends to decay 
into the second excited state. This is surprising; one might instead 
have expected the tendency to decay into the ground state.
Since the energy of the second excited state is higher than that of 
the first excited state, it would seem physically reasonable that this decay
cannot happen spontaneously; this gives a simple explanation for stability.

We point out that these stability results are only valid
for {\em{weak}} coupling, i.e., for small (scaled) mass $m$.
If $m$ comes into the region where the mass spectrum of Figure 
\ref{bind} starts to have the form of a spiral curve, the numerical
behavior of the linearized equations becomes much more difficult to 
analyze and does no longer allow simple conclusions. It is thus very 
helpful to study the stability in this regime via a different
method, which we will now describe.

The existence of the spiral curve (c.f.\ Figure \ref{spiral}), 
enables us to obtain information regarding the stability properties 
of these solutions, using Conley Index theory (see \cite[Part IV]{14a}).
Indeed, as the figure shows, if $m>m_1$, there are no solutions while 
at $m=m_1$, the solution $P_1$ suddenly appears. This solution is 
``degenerate'', and has Conley index $\overline{0}$ (the homotopy 
type of a one-point, pointed space). For $m<m_1$, the solution $P_1$
bifurcates into two solutions $Q_1$ and $Q_2$. The solution $Q_1$, 
being a ``continuation'' of the stable solution $Q_0$ ($m$ is near $0$), 
must also be stable; this follows from Conley's Continuation Theorem
\cite[Thm. 23.31]{14a}. In fact, since $Q_0$ being stable implies that 
the Conley index of $Q_0$ is $\Sigma^0$, (the pointed zero sphere), 
the continuation theorem implies that the Conley index of $Q_1$ must 
also be $\Sigma^0$, and thus $Q_1$ is also a stable solution.
(Moreover, the same argument shows that all points on the curve 
between $0$ and $P_1$ correspond to stable solutions; this can also be 
checked numerically). Since the Conley indices of $Q_1$ and $Q_2$ must 
``cancel'' (i.e., the index of $Q_1$ and $Q_2$ taken together must 
be $\overline{0}$), this implies that $Q_2$ cannot be stable, and in 
fact, the index of $Q_2$ is $\Sigma^1$, the pointed 1-sphere. (In 
fact, all points on the curve between $P_1$ and $P_2$ correspond to 
unstable solutions; one can also check this numerically.)
Similar reasoning can be applied to solutions near $P_2$, $P_3$, 
\ldots, and so on. (We remark that only $m$, and {\em{not}} $\omega$,
can serve as a bifurcation parameter; we show this in Appendix 
\ref{bfp}.)

These general Conley Index techniques also enable us to show that for 
each $n$, these spiral curves must tend, as $\alpha_1 \rightarrow 
\infty$, to a limiting configuration $\Gamma_n$, which is either a 
single point, or is a ``limit cycle'' $S^1$; i.e.\ a topological 
circle (we assume, as the numerics indicate, that the curve 
``spirals inwards''). In fact, were this not the case, then for each value of 
$m$ between $0$ and $m_1$, the corresponding solution 
set would form an ``isolated invariant set'' (c.f. \cite{14a}), and 
so, again by Conley's Continuation Theorem, their Conley indices 
would all be the same. However, for $m$ near $0$, the index of 
the isolated invariant set is $\Sigma^0$, while for $m$ near 
$m_1$, the corresponding isolated invariant set has index 
$\overline{0}$. Since these two indices are different, we have a 
contradiction.

It follows from this last result that for a point
$(\hat{m}, \hat{m}-\hat{\omega})$ on $\Gamma_n$, there are an infinite number of 
solutions with $m=\hat{m}$, as well as an infinite number of solutions 
with $m-\omega=\hat{m}-\hat{\omega}$. For parameter points not meeting
$\Gamma_n$, there are at most a finite number of solutions.

\appendix
\section{Appendix: Justification of the Ansatz for the Spinor Dependence}
\label{asd}\setcounter{equation}{0}
In this section, we consider the general form of the spinors 
in static, spherically symmetric systems and derive the Einstein-Dirac
equations for these systems.
This analysis will clarify the ansatz' (\ref{0}) and (\ref{0z}) for the 
wave functions, which was made in Section \ref{sec4} without a 
detailed explanation.

The Dirac wave functions $\Psi_1$, $\Psi_2$ of a general two-fermion 
system can be written in the form
\begin{equation}
	\Psi_a(\vec{x}, t) \;=\; A(\vec{x}, t) \:e_a \;\;\;,\spc a=1,2\spc .
	\label{A1}
\end{equation}
where $A^\alpha_a=\Psi^\alpha_a$ is a $(4 \times 2)$-matrix and 
where $(e_a)$ again denotes the standard basis of the two-component 
Pauli spinors. The system being {\em{static}} means that the time 
dependence of $A$ has the form of a plane wave,
\[ A(\vec{x}, t) \;=\; e^{-i \omega t} \:A(\vec{x}) \spc . \]
The simplest way to characterize the {\em{spherical symmetry}} of the 
spinors is to demand that the angular dependence of $A$ is described 
only by the submatrices $\1$ and $\sigma^r$; i.e.
\begin{equation}
	A(\vec{x}) \;=\; \left( \begin{array}{c} v_1(r) \:\1 \:+\: v_2(r) 
	\: \sigma^r \\ v_3(r) \:\1 \:+\: v_4(r) \:\sigma^r
	\end{array} \right)
	\label{A2}
\end{equation}
with complex functions $v_1,\ldots,v_4$. This form of $A$ can be 
derived if one requires that the total angular momentum is zero and 
that all the expectation values $\bra {\cal{O}} \ket=\sum_{a=1}^2 
\overline{\Psi_a} {\cal{O}} \Psi_a$ of the spin matrices are 
spherically symmetric. In a simplified argument, this form can be 
understood directly from the fact that the presence of any matrices 
$\sigma^\vartheta$, $\sigma^\varphi$ would destroy the radial symmetry 
in (\ref{A2}).

The ansatz (\ref{A1}), (\ref{A2}) for the wave functions is a linear 
combination of (\ref{0}) and (\ref{0z}). From this, we immediately 
obtain the corresponding Dirac equations. Namely, the complex 
two-spinors $\Phi=(\Phi_1, \Phi_2)$ and $\Xi=(\Xi_1, \Xi_2)$ with
\begin{eqnarray*}
	\Phi_1 & = & r \:T^{-\frac{1}{2}} \:v_1 \;\;\;\;\;\;\;,\spc
	\Xi_1 \;=\; r \:T^{-\frac{1}{2}} \:v_2 \\
	\Phi_2 & = & -i r \:T^{-\frac{1}{2}} \:v_4 \;\;\;,\spc
	\Xi_2 \;=\; -i r \:T^{-\frac{1}{2}} \:v_3
\end{eqnarray*}
satisfy, (in analogy to (\ref{3})), the equations
\begin{eqnarray}
\sqrt{A} \: \Phi^\prime &=& \left[ \omega T
	\left( \begin{array}{cc} 0 & -1 \\ 1 & 0 \end{array} \right)
	\:+\: \frac{1}{r} \:\left( \begin{array}{cc} 1 & 0 \\
    0 & -1 \end{array} \right) \:-\: m\:
    \left( \begin{array}{cc} 0 & 1 \\
    1 & 0 \end{array} \right) \right] \Phi \label{A3} \\
\sqrt{A} \: \Xi^\prime &=& \left[ \omega T
	\left( \begin{array}{cc} 0 & -1 \\ 1 & 0 \end{array} \right)
	\:-\: \frac{1}{r} \:\left( \begin{array}{cc} 1 & 0 \\
    0 & -1 \end{array} \right) \:-\: m\:
    \left( \begin{array}{cc} 0 & 1 \\
    1 & 0 \end{array} \right) \right] \Xi \spc .
	\label{A4}
\end{eqnarray}
In Section \ref{sec4}, we argued that the reality of the coefficients 
in (\ref{3}) allows us to choose real spinors. This procedure 
simplified the Dirac equations considerably, but it might be too restrictive
to describe the general solution of the Einstein-Dirac equations. In 
order to analyze the situation more carefully, we first note that the 
function
\[ F(r) \;:=\; \Phi(r)^* \left( \begin{array}{cc}
0 & -i \\ i & 0 \end{array} \right) \Phi(r) \]
is independent of $r$, as is obvious from (\ref{A3}).
The boundary conditions at the origin, $\Phi_1(0)=0=\Phi_2(0)$,
imply that $F$ must vanish identically, and thus the product 
$\overline{\Phi_1} \Phi_2$ is real. This means that $\Phi_1$ and
$\Phi_2$ must be real up to a common phase factor, i.e.
\begin{equation}
\Phi_1(r) \;=\; e^{i \alpha} \:f_1(r) \;\;\;,\spc \Phi_2(r) \;=\; e^{i 
\alpha} \: f_2(r)
	\label{A8}
\end{equation}
with real functions $f_1, f_2$. Again as a consequence of the Dirac 
equation (\ref{A3}), the phase $\alpha$ is independent of $r$. A similar
argument applies to $\Xi$ and gives
\begin{equation}
	\Xi_1 \;=\; e^{i \beta} \:g_1(r) \;\;\;,\spc \Xi_2 \;=\; e^{i 
	\beta} \:g_2(r)
	\label{A9}
\end{equation}
with real functions $g_1, g_2$ and $\beta \in \R$.

We shall now compute the energy-momentum tensor. First of 
all, the spherical symmetry of the spinors implies that the 
off-diagonal components $T^t_\vartheta$, $T^t_\varphi$, $T^r_\vartheta$,
$T^r_\varphi$, $T^\vartheta_\varphi$ vanish and that 
$T^\vartheta_\vartheta = T^\varphi_\varphi$. Thus we must only 
consider $T^t_r$ and the diagonal components $T^t_t$, $T^r_r$, 
$T^\vartheta_\vartheta$ of the energy-momentum tensor.
As a first step, we verify that the contribution 
(\ref{42a}) of the variation of $B$ vanishes: According to 
(\ref{2.16c}), the trace in (\ref{42a}) is zero if $\delta g_{mk}$ is 
diagonal. Thus we must only consider $T^t_r$, and we may therefore
assume that the indices $m, k$ are either $m=t$, $k=r$ or $m=r$, $k=t$.
Furthermore, the spherical symmetry (\ref{A2}) implies that the 
expectation of the pseudovector $\gamma^5 G^p$ only has a component 
in the time and radial directions,
\[ \sum_{a=1}^2 \overline{\Psi_a} \:\gamma^5 G^p\:\Psi_a \;=\; 0 \spc
{\mbox{for $p=\varphi$ or $p=\vartheta$}} \spc . \]
We conclude that both indices $m$ and $p$ in (\ref{42a}) must be equal to
either $r$ or $t$, and the antisymmetry of the $\epsilon$-tensor implies that 
the remaining indices $j, n$ must coincide either with $\vartheta$ or 
$\varphi$. Thus we must only consider the trace in (\ref{42a}) for 
the combination
\begin{equation}
	\Tr \left( G^k \: (\partial_\vartheta G_\varphi - 
	\partial_\varphi G_\vartheta) \right) \spc .
	\label{A3a}
\end{equation}
But, according to (\ref{2.13aa}), (\ref{2.13b}) and (\ref{29a}), we have 
$\partial_\vartheta G_\varphi = \partial_\varphi G_\vartheta$, so
that (\ref{A3a}) vanishes.

We conclude that the energy-momentum tensor is again given by (\ref{4.3}).
In order to compute the trace in (\ref{4.3}), it is useful to
first notice that, if we write the Dirac matrices as $(2 \times 2)$ 
block matrices, then $G^t$ is diagonal with entries proportional to the
identity, whereas $G^r$ is off-diagonal with 
submatrices which are multiples of $\sigma^r$. This implies that the mixed 
contribution (i.e., the contribution proportional to $\overline{\Phi} 
\Xi$ or $\overline{\Xi} \Phi$) to $T^t_t$ and $T^r_r$ vanish.
Using the explicit form of $G^\vartheta$ together with (\ref{d}), we 
conclude that the mixed contribution also vanishes in 
$T^\vartheta_\vartheta$. Thus the energy-momentum tensor of the 
system is simply the sum of the energy-momentum tensors corresponding 
to the spinors $\Phi$ and $\Xi$. As a consequence, the constant phase 
factors in (\ref{A8}), (\ref{A9}) are irrelevant; we can without loss 
of generality assume that $\Phi$ and $\Xi$ are real. Using (\ref{c}), we 
end up with the formulas
\begin{eqnarray}
	T^i_j & = & T^i_j[\Phi] \:+\: T^i_j[\Xi] \spc {\mbox{with}} \\
	\label{A10}  \\
T^i_j[\Phi] & = & r^{-2} \;{\mbox{diag }} \left(
	2 \omega \:T^2\: |\Phi|^2, \;
	-2 \omega \:T^2\: |\Phi|^2 + 4 T \: r^{-1} \:
	\Phi_1 \Phi_2 \:+\: 2 m \:T \:(\Phi_1^2 - \Phi_2^2),
	\right. \nonumber \\
&&\hspace*{3cm} \left. -2 T \:r^{-1} \:\Phi_1 \Phi_2,\;
	-2 T \:r^{-1} \:\Phi_1 \Phi_2 \right)
	\label{A11} \\
T^i_j[\Xi] & = & r^{-2} \;{\mbox{diag }} \left(
	2 \omega \:T^2\: |\Xi|^2, \;
	-2 \omega \:T^2\: |\Xi|^2 - 4 T \: r^{-1} \:
	\Xi_1 \Xi_2 \:+\: 2 m \:T \:(\Xi_1^2 - \Xi_2^2),
	\right. \nonumber \\
&&\hspace*{3cm} \left. 2 T \:r^{-1} \:\Xi_1 \Xi_2,\;
	2 T \:r^{-1} \:\Xi_1 \Xi_2 \right) \spc .
	\label{A12}
\end{eqnarray}
Thus the Einstein-Dirac equations take the form (\ref{A3}),(\ref{A4}) 
and
\[ G^i_j \;=\; -8 \pi \left( T^i_j[\Phi] + T^i_j[\Xi] \right) \]
with $T^i_j$ given by (\ref{A11}), (\ref{A12}). This is a 
generalization of the system (\ref{dirac1})-(\ref{Teq}). 
It is quite remarkable that the energy-momentum tensor is just the sum 
of $T^i_j[\Phi]$ and $T^i_j[\Xi]$. Similar to (\ref{norm}), the normalization 
condition for the wave functions takes the form
\begin{equation}
	\int_0^\infty \left(|\Phi|^2 + |\Xi|^2\right)
	\:\frac{T}{\sqrt{A}} \; dr \;=\; \frac{1}{4 \pi} \spc .
	\label{A13}
\end{equation}

We now qualitatively describe how the solutions of this generalized 
system can be constructed and how we recover the solutions of 
Section \ref{sec9}.
The scaling technique of Section \ref{sec8} can again be applied 
and consequently we can omit the conditions $T(\infty)=1$ and (\ref{A13}) 
if we instead set $T(0)=1=m$. Then the solutions are characterized by 
the three parameters $\omega, \Phi_1^\prime(0), \Xi_2^\prime(0)$
(notice that a Taylor expansion around $r=0$ yields, in analogy to (\ref{ic2}),
the constraints $\Phi_2'(0)=0=\Xi_1'(0)$).
Compared to the situation in Section \ref{sec9}, we thus have one 
additional continuous parameter to describe the solution. At first 
sight, this might seem to imply that we can now construct, for given $\omega$, 
a continuous one-parameter family of solutions. Then our ansatz'
(\ref{0}) and (\ref{0z}) would just correspond to two special points 
of this continuum of solutions, and it would become unsatisfying that 
we just picked these two special solutions for the discussion of the 
mass spectrum and the stability.
However, the additional free parameter is illusory due to the 
fact that we also have one additional constraint at infinity.
Namely, we must (for given $\omega$) adjust $\Phi_1^\prime(0)$ and 
$\Xi_2^\prime(0)$ in such a way that both the $(\Phi_1, \Phi_2)$ and
the $(\Xi_1, \Xi_2)$ curves tend to zero for large $r$. For generic $\omega$,
these conditions will only 
be satisfied for a discrete set of initial values $(\Phi_1^\prime(0), 
\Xi_2^\prime(0))$. The choices $(\Phi_1^\prime(0)=0,\: \Xi_2^\prime(0))$
and $(\Phi_1^\prime(0), \:\Xi_2^\prime(0)=0)$ are both allowable; they 
correspond to the solutions constructed in Section \ref{sec9}.
After scaling, this shows that for generic $m$, the 
Einstein-Dirac equations only admit a discrete number of solutions.

We note that these considerations do not rule out the possibility that the
general ansatz for the spinors might lead to some additional solutions. We 
did not study the general 
equations systematically and can only say qualitatively that it seems 
difficult to arrange that there are simultaneous normalizable 
solutions of (\ref{A3}) and (\ref{A4}).
The existence of solutions of this type, however, remains an open question.

\section{Appendix: Justification of Using $m$ as the Bifurcation Parameter}
\label{bfp}\setcounter{equation}{0}
We show in this section, first that $\omega$ is unsuitable as a 
bifurcation parameter, and second that $m$ can serve as a bifurcation 
parameter. In Conley Index theory, a parameter can only be used as a 
bifurcation parameter if it remains well-defined and fixed when 
perturbations of the solutions are considered. The basic reason why 
$m$ and $\omega$ must be treated differently can already be understood 
from the general form of the Dirac equation in (\ref{1.1}). The mass 
$m$ enters as an a-priori given parameter into the Dirac equation, 
whereas the energy $\omega$ of the fermion is only determined by the 
solution $\Psi$. This means that if we consider a variation of a 
solution, $m$ can be considered as a fixed parameter, while $\omega$ 
will in general change. If the perturbation of the solution is not 
static, the energy of the fermion will in general become 
time-dependent, and $\omega$ will no longer be a well-defined parameter.

Although being correct in principle, this argument is too simple and 
not fully convincing. First of all, the situation becomes more 
complicated by our scaling technique, which also changes $m$ and 
thus makes it impossible to consider the mass as a fixed parameter 
throughout. Furthermore, $\omega$ is uniquely determined by the 
solutions $(\alpha, \beta, A, T)$ of (\ref{dirac1})-(\ref{Teq}).
It enters as a parameter into the linearized equations in a similar 
way as $m$ does, and it is not obvious from these equations why the 
two parameters $m$ and $\omega$ should play such different roles for 
stability considerations. Therefore we will show in detail that 
solutions of the linearized equations do not determine $\omega$, 
whereas $m$ is still a well-defined parameter.

In order to show that $\omega$ is not well-defined, we generalize the 
ansatz of Section \ref{sec10} in the way that we also allow $\omega$ 
to be time-dependent. In analogy to (\ref{F})-(\ref{G}) and (\ref{D}),(\ref{E}),
we consider a perturbation of $\omega$ of the form
\begin{equation}
	\omega(t) \;=\; \omega \:+\: \epsilon \:\omega_1 \:e^{\kappa t} \spc .
	\label{bo1}
\end{equation}
Since $\omega$ represents a frequency, i.e.\ the time-derivative of a phase,
the correct generalization of equation (\ref{F2}) is to replace the 
phase factor $e^{-i \omega t}$ by
\[ \exp \left( -i \int_0^t \omega(s) \:ds \right) \spc . \]
The ansatz for $a_j$, $b_j$, $A_1$, $T_1$ then remains the same
as before, given by (\ref{F})-(\ref{G}) and (\ref{D}),(\ref{E}).
Thus the spinors $z_1$, $z_2$ are given by
\begin{eqnarray*}
z_1(r,t) & = & e^{-i \int_0^t \omega(s) \:ds} \left[ \alpha(r) 
	\:+\: \varepsilon (a_1(r,t)+i \:a_2(r,t) \right] \\
z_2(r,t) & = & e^{-i \int_0^t \omega(s) \:ds} \left[ \beta(r) 
	\:+\: \varepsilon (b_1(r,t)+i \:b_2(r,t) \right] \spc .
\end{eqnarray*}
Into these equations, we substitute (\ref{bo1}) and consider only the 
first-order terms in $\varepsilon$. This gives
\begin{eqnarray*}
z_1(r,t) & = & e^{-i \omega t} \left(1 - i \varepsilon \int_0^t 
	\omega_1 \:e^{\kappa s} \:ds \right) \left[ \alpha(r) 
	\:+\: \varepsilon (a_1(r,t)+i \:a_2(r,t) \right] \\
&=& e^{-i \omega t} \left[ \alpha(r) \:-\: i \varepsilon 
\:\frac{\omega_1}{\kappa} \:(e^{\kappa t}-1) \:\alpha(r)
	\:+\: \varepsilon (a_1(r,t)+i \:a_2(r,t) \right] \\
&=& e^{-i \omega t} \left[ (1+ i \varepsilon \:\frac{\omega_1}{\kappa}) 
\:\alpha(r) \:+\: \varepsilon (a_1(r,t)+i \:a_2(r,t) 
\:-\: i \varepsilon 
\:\frac{\omega_1}{\kappa} \:e^{\kappa t} \:\alpha(r)\right] \\
&=& e^{-i (\omega t - \varepsilon \:\frac{\omega_1}{\kappa})}
\left[ \alpha(r) \:+\: \varepsilon (a_1(r,t)+i \:a_2(r,t) 
\:-\: i \varepsilon \:\frac{\omega_1}{\kappa} \:e^{\kappa t} \:\alpha(r)\right]
\end{eqnarray*}
with a similar expression for $z_2(r,t)$. This looks quite similar to 
the original ansatz (\ref{F2}) and (\ref{F})-(\ref{G}) except for two 
differences; namely, there is here an additional constant phase 
factor $\exp(i \varepsilon \:\frac{\omega_1}{\kappa})$, together with 
the term
\begin{equation}
	-i \varepsilon \:\frac{\omega_1}{\kappa}\:e^{\kappa t}\:\alpha(r) 
	\spc .
	\label{bo2}
\end{equation}
The phase factor $\exp(i \varepsilon \:\frac{\omega_1}{\kappa})$ plays 
no role in our analysis, since it falls out of all the equations 
(notice that the energy-momentum tensor (\ref{4.3}) contains only terms of
the form $\overline{\Psi} \cdots \Psi$). What is interesting about the term 
(\ref{bo2}), however, is that its time-dependence is again of the form 
$e^{\kappa t}$. It is thus consistent with our ansatz (\ref{D}), and 
corresponds to the transformation
\begin{equation}
	a_2(r) \;\rightarrow\; a_2(r) \:-\: \varepsilon 
	\:\frac{\omega_1}{\kappa} \:\alpha(r) \spc ,
	\label{bo3}
\end{equation}
and similarly
\begin{equation}
	b_2(r) \;\rightarrow\; b_2(r) \:-\: \varepsilon 
	\:\frac{\omega_1}{\kappa} \:\beta(r) \spc .
	\label{bo4}
\end{equation}
Thus the more general ansatz for $a_j, b_j, A_1, T_1$ whereby $\omega$ 
is replaced by (\ref{bo1}) is the same as the original ansatz for 
these quantities if we transform $a_2$ and $b_2$ according to 
(\ref{bo3}) and (\ref{bo4}). Conversely, we may obtain an arbitrary 
time dependence in $\omega$, of the form (\ref{bo1}), merely by 
transforming $a_2$ and $b_2$ according to (\ref{bo3}) and (\ref{bo4}).
This means that a solution of the linearized equations only 
determines $\omega$ up to linear time-dependent perturbations of the 
form (\ref{bo1}). Thus $\omega$ is not a well-defined 
parameter\footnote{It is interesting to notice that the contribution 
to $(a_2, b_2)$ proportional to $(\alpha, \beta)$ which occurs in the 
transformation (\ref{bo3}),(\ref{bo4}) played an important role in our
numerics. Namely, we saw in Section \ref{sec10} that the $(a_2, 
b_2)$-plot looks very similar to the $(\alpha, \beta)$-plot (see 
Figure \ref{stab1}), which shows that this contribution is actually 
dominant for small $\kappa$. It caused problems in the numerics and 
forced us to introduce the new variables $\hat{a}_2$, $\hat{b}_2$ (see 
Figure \ref{stab2}). According to (\ref{bo3}),(\ref{bo4}), we can now 
understand the dominant contribution to $(a_2, b_2)$ in Figure 
\ref{stab1} as describing a time-dependent perturbation of $\omega$ of
the form (\ref{bo1}).}.

For the parameter $m$, however, the situation is completely 
differently. Namely if $\Psi$ is a solution of the time-dependent 
Dirac equation $G \Psi = m \Psi$, then we see from (\ref{nc}) that 
$(G \Psi \:|\: G \Psi) = m^2$. But as we noted earlier, (after 
(\ref{C})), this relation is also valid for the linearized equations. 
That is, $m$ is a well-defined parameter for the linearized equations.

\clearpage
\begin{figure}[thbp]
	\epsfxsize=18cm
	\centerline{\epsfbox{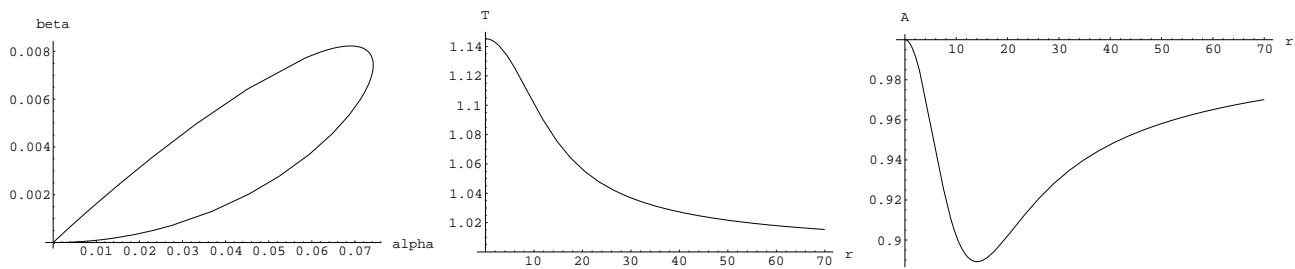}}
	\caption{The ground state for $m=1$, $\alpha_1=0.02$; scaled 
	parameter values: $m=0.5340$, $\omega=0.4994$}
	\label{ground}
\end{figure}
\begin{figure}[thbp]
	\epsfxsize=18cm
	\centerline{\epsfbox{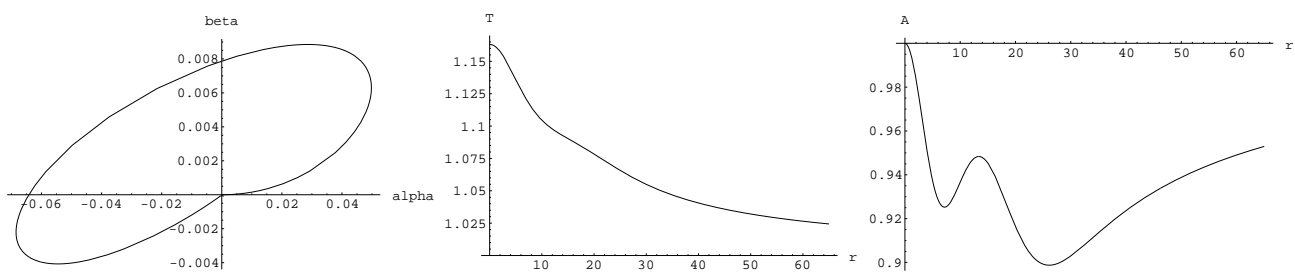}}
	\caption{The first excited state for $m=1$, $\alpha_1=0.02$; scaled 
	parameter values: $m=0.7779$, $\omega=0.7326$}
	\label{first}
\end{figure}
\begin{figure}[thbp]
	\epsfxsize=18cm
	\centerline{\epsfbox{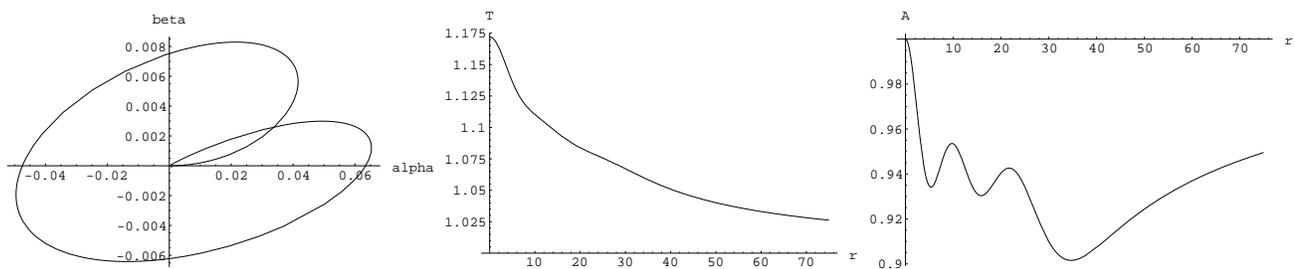}}
	\caption{The second excited state for $m=1$, $\alpha_1=0.02$; scaled 
	parameter values: $m=0.9616$, $\omega=0.9080$}
	\label{second}
\end{figure}
\begin{figure}[thbp]
	\epsfxsize=18cm
	\centerline{\epsfbox{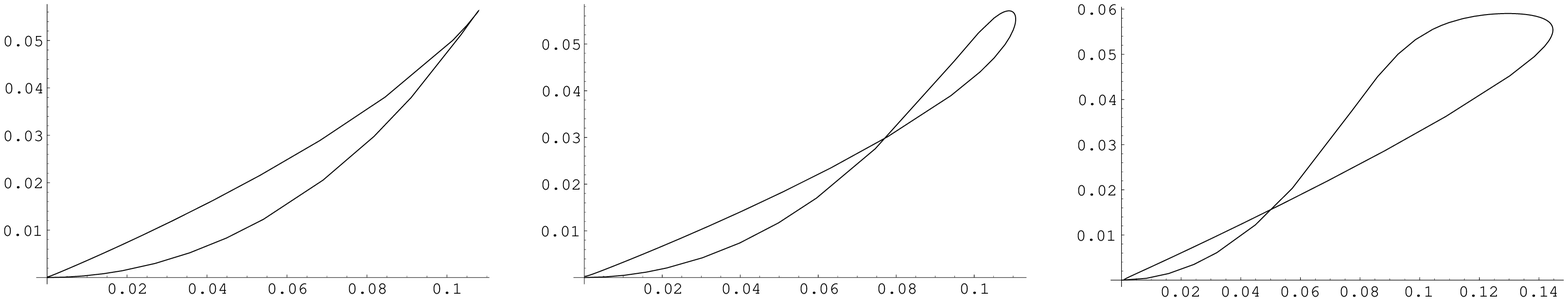}}
	\caption{$\alpha$-$\beta$-plot for the ground states at parameter 
	values $m=1$ and $\alpha_1=0.31, 0.35, 0.45$}
	\label{loop}
\end{figure}
\begin{figure}[thbp]
	\epsfxsize=18cm
	\centerline{\epsfbox{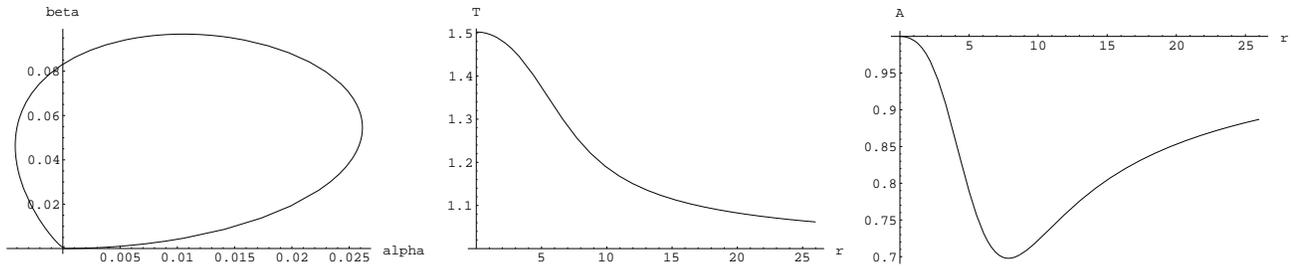}}
	\caption{The lowest state for $m=-1$, $\alpha_1=0.02$; scaled 
	parameter values: $m=-0.7567$, $\omega=0.6302$}
	\label{nfirst}
\end{figure}
\begin{figure}[thbp]
	\epsfxsize=18cm
	\centerline{\epsfbox{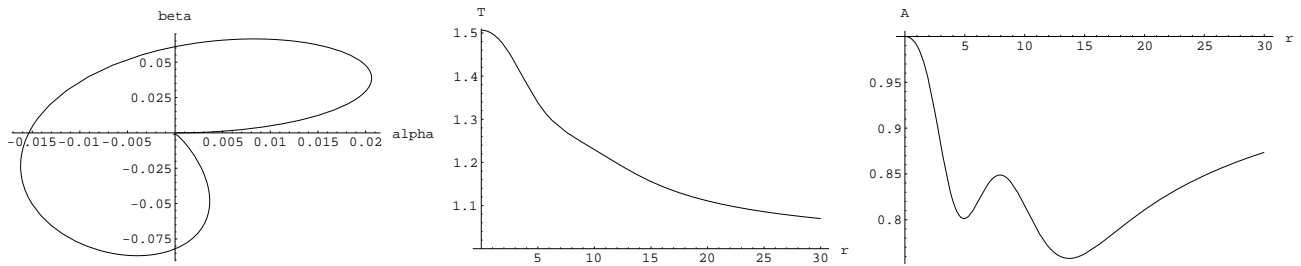}}
	\caption{The next excited state for $m=-1$, $\alpha_1=0.02$; scaled 
	parameter values: $m=-0.9742$, $\omega=0.8391$}
	\label{nsecond}
\end{figure}
\begin{figure}[thbp]
	\epsfxsize=13cm
	\centerline{\epsfbox{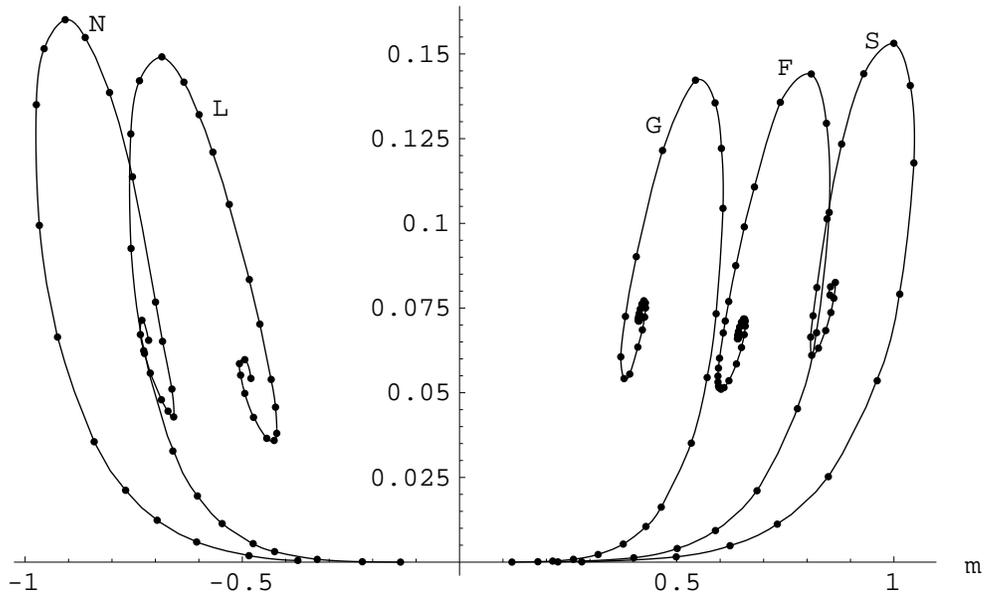}}
	\caption{Binding Energy $|m|-\omega$ of the Fermions}
	\label{bind}
\end{figure}
\begin{figure}[thbp]
	\epsfxsize=15cm
	\centerline{\epsfbox{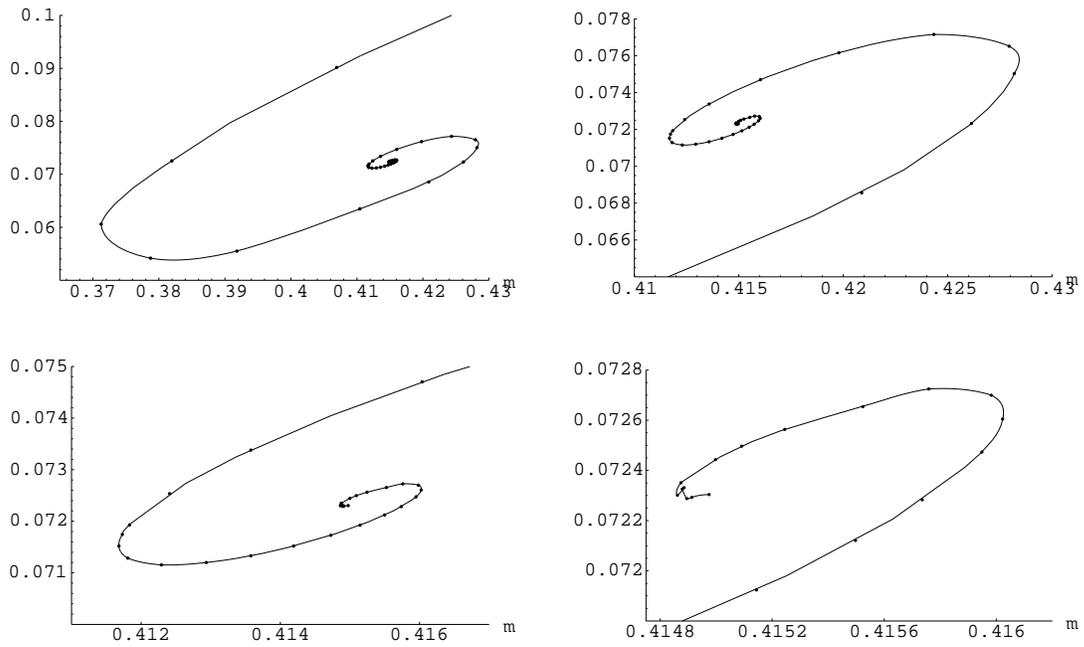}}
	\caption{Binding Energy $|m|-\omega$  of the Ground State, Detailed 
	Pictures}
	\label{zoom}
\end{figure}
\begin{figure}[thbp]
	\epsfxsize=13cm
	\centerline{\epsfbox{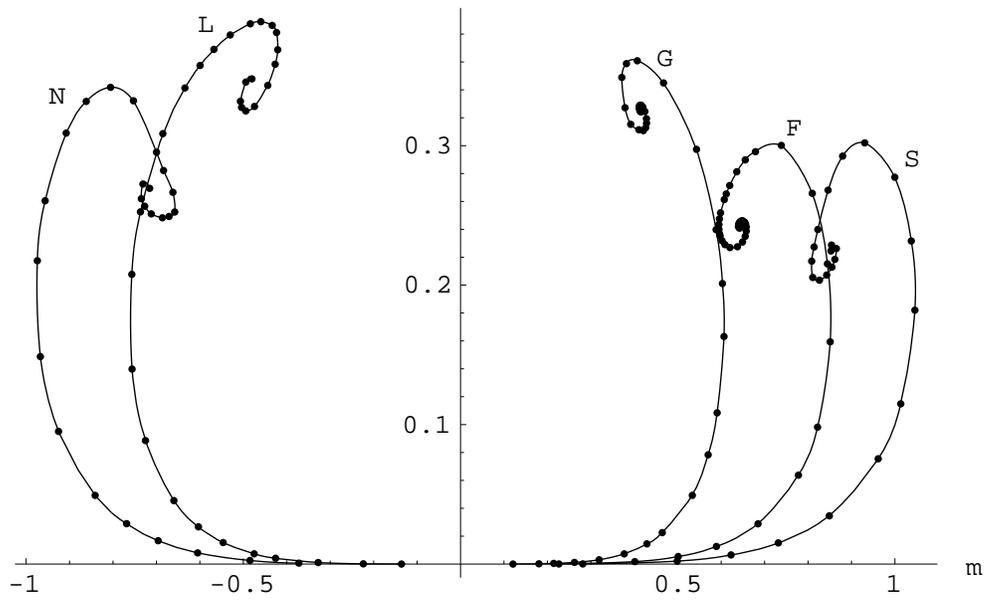}}
	\caption{Total Energy $\rho-2 \omega$ of the Gravitational Field}
	\label{adm}
\end{figure}
\begin{figure}[thbp]
	\epsfxsize=13cm
	\centerline{\epsfbox{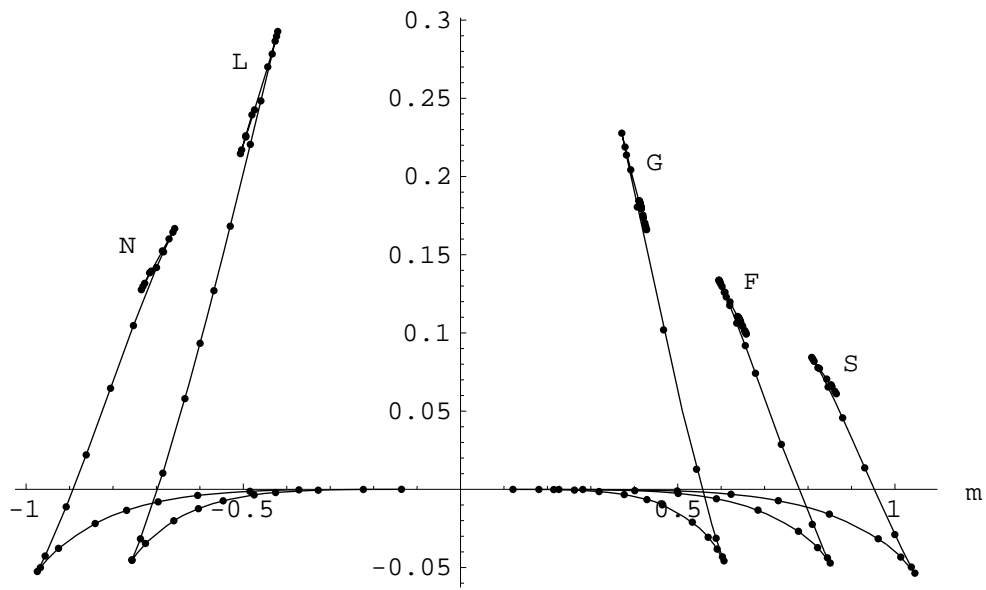}}
	\caption{Total Binding Energy $\rho-2 |m|$}
	\label{tm}
\end{figure}
\clearpage
\begin{figure}[thbp]
	\epsfxsize=15cm
	\centerline{\epsfbox{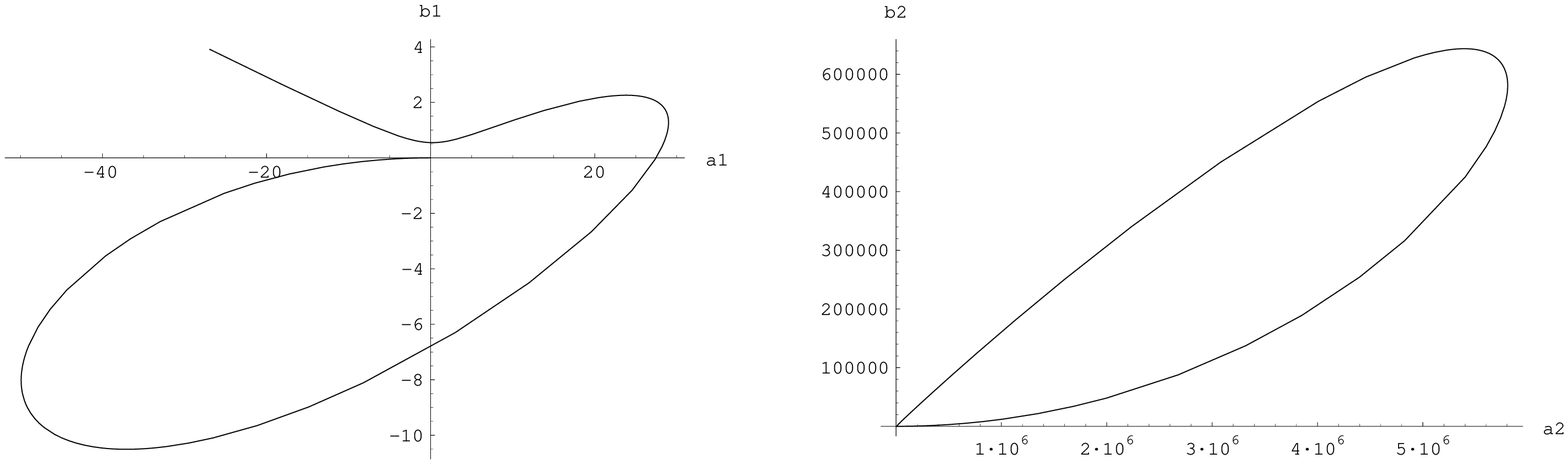}}
	\caption{Perturbation of the ground state for $\kappa=10^{-6}$,
	variables $(a_1, b_1, a_2, b_2)$}
	\label{stab1}
\end{figure}
\begin{figure}[thbp]
	\epsfxsize=15cm
	\centerline{\epsfbox{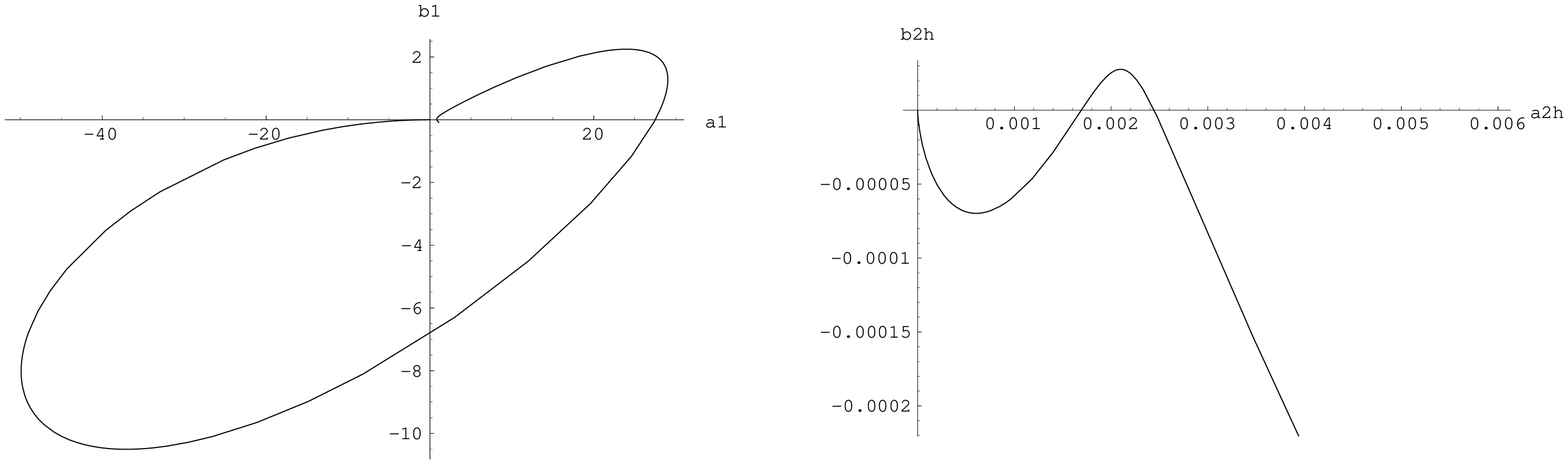}}
	\caption{Perturbation of the ground state for $\kappa=10^{-6}$,
	variables $(a_1, b_1, \hat{a}_2, \hat{b}_2)$}
	\label{stab2}
\end{figure}
\begin{figure}[thbp]
	\epsfxsize=15cm
	\centerline{\epsfbox{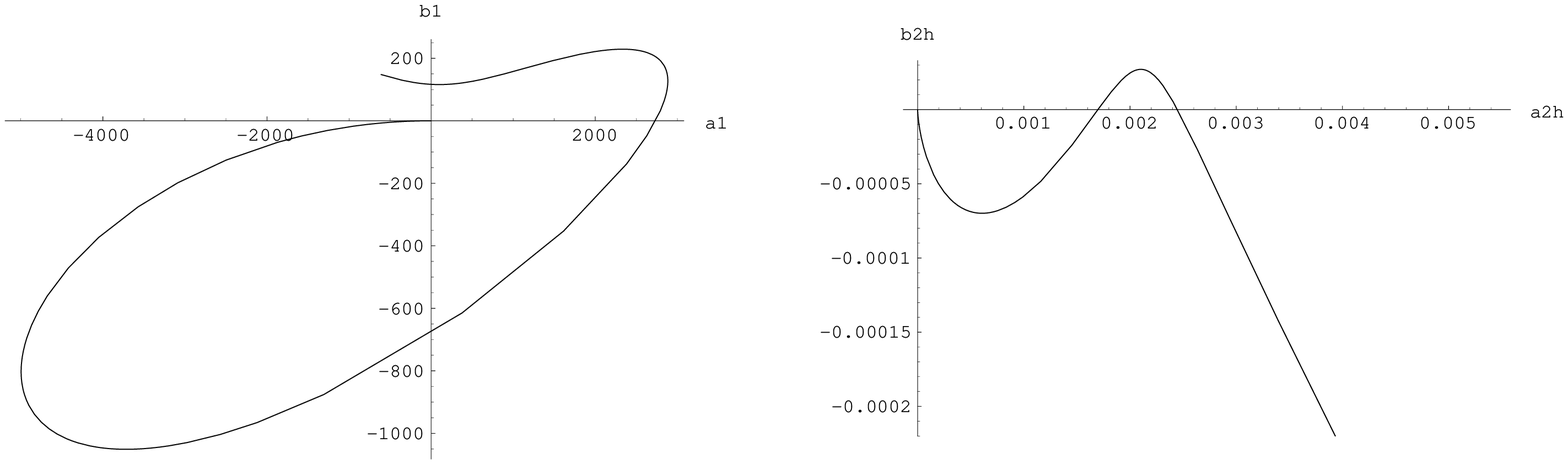}}
	\caption{Perturbation of the ground state for $\kappa=10^{-8}$}
	\label{sg1}
\end{figure}
\begin{figure}[thbp]
	\epsfxsize=15cm
	\centerline{\epsfbox{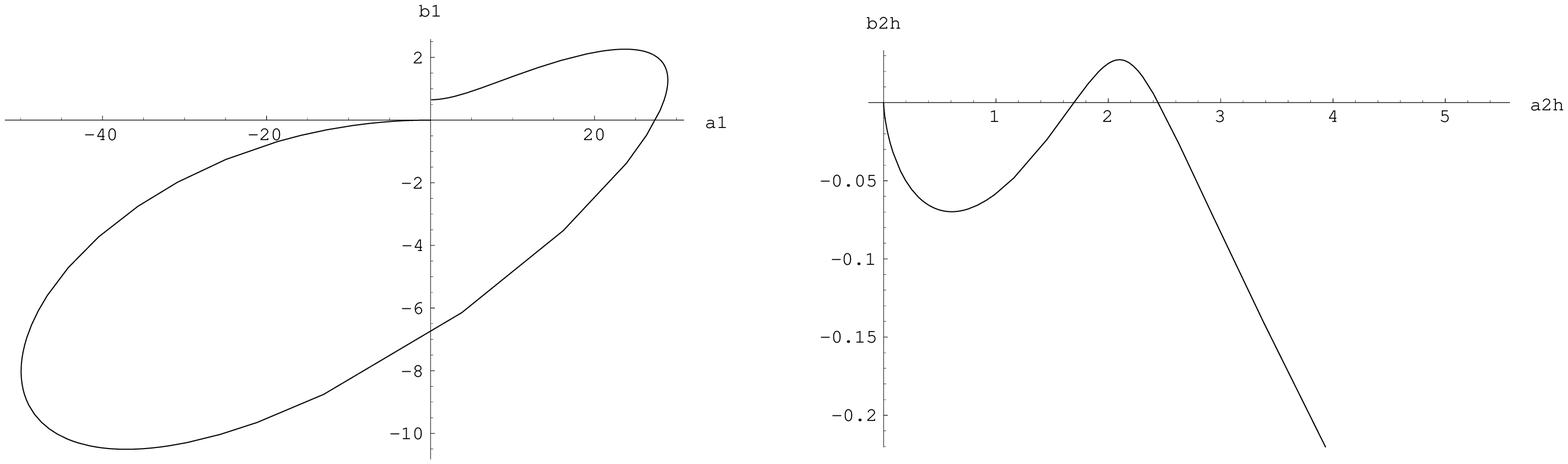}}
	\caption{Perturbation of the ground state for $\kappa=10^{-3}$}
	\label{sg2}
\end{figure}
\begin{figure}[thbp]
	\epsfxsize=15cm
	\centerline{\epsfbox{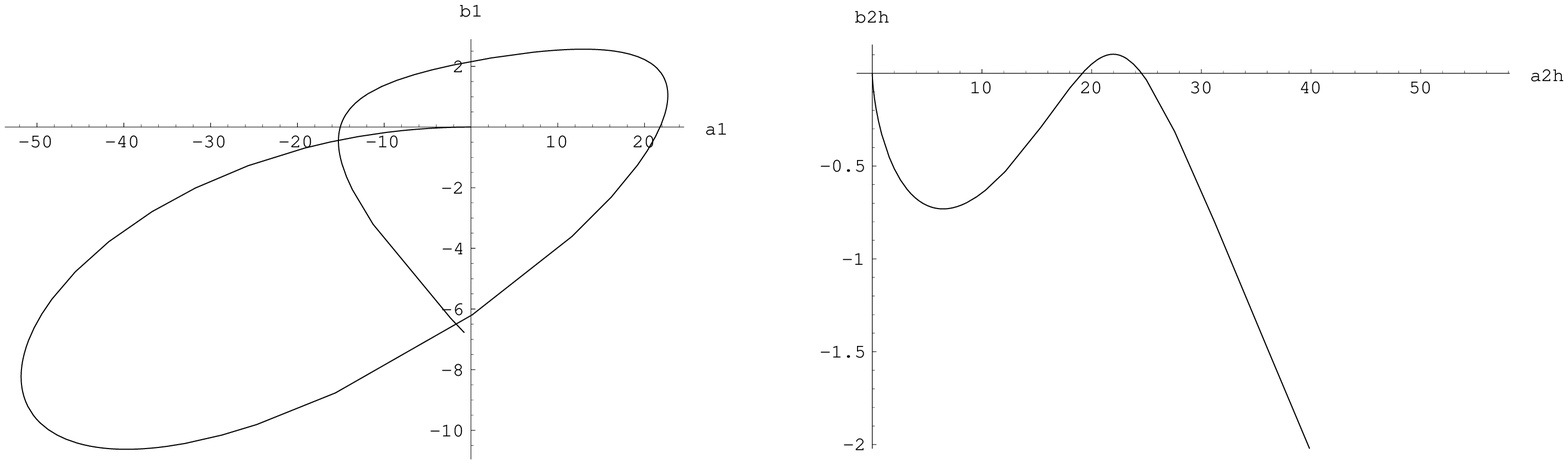}}
	\caption{Perturbation of the ground state for $\kappa=0.01$}
	\label{sg3}
\end{figure}
\begin{figure}[thbp]
	\epsfxsize=15cm
	\centerline{\epsfbox{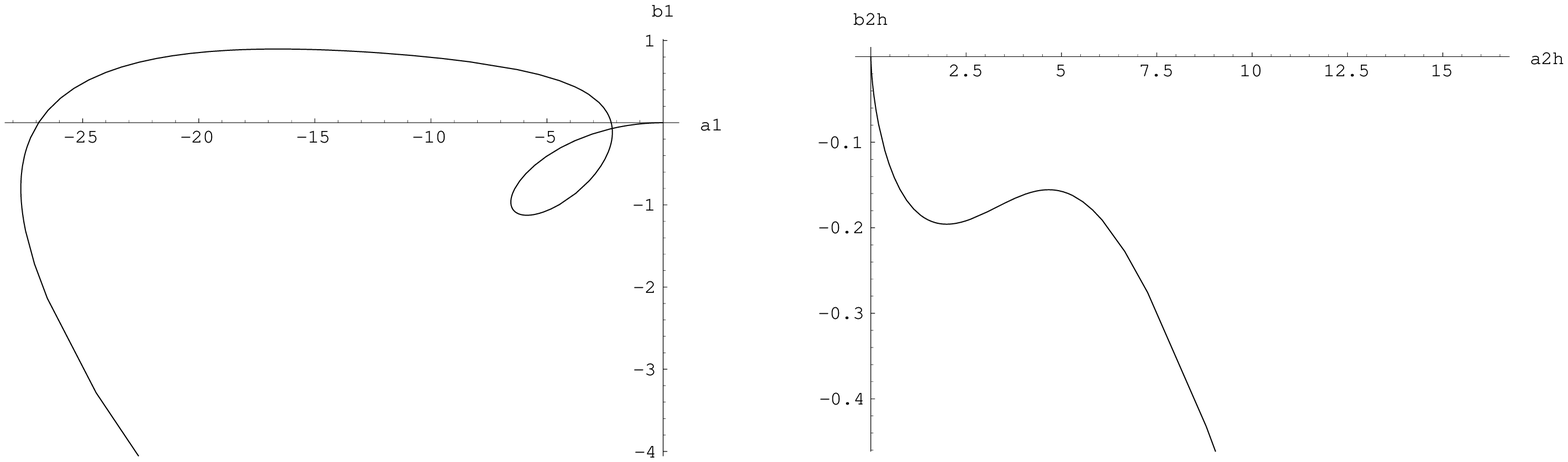}}
	\caption{Perturbation of the ground state for $\kappa=0.02$}
	\label{sg4}
\end{figure}
\begin{figure}[thbp]
	\epsfxsize=15cm
	\centerline{\epsfbox{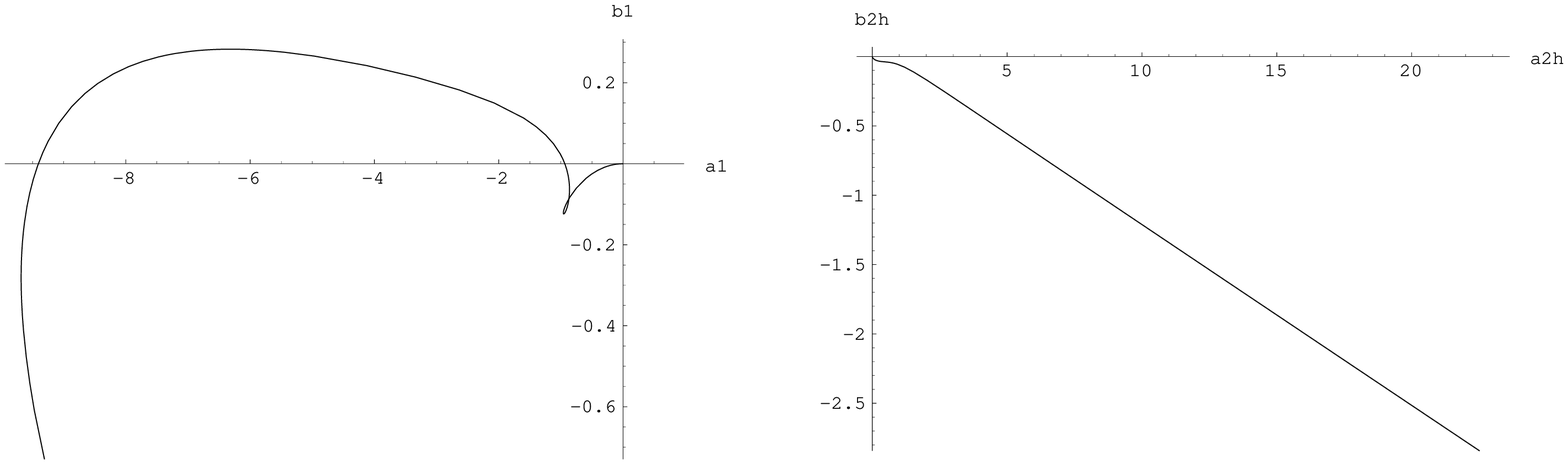}}
	\caption{Perturbation of the ground state for $\kappa=0.023$}
	\label{sg5}
\end{figure}
\begin{figure}[thbp]
	\epsfxsize=15cm
	\centerline{\epsfbox{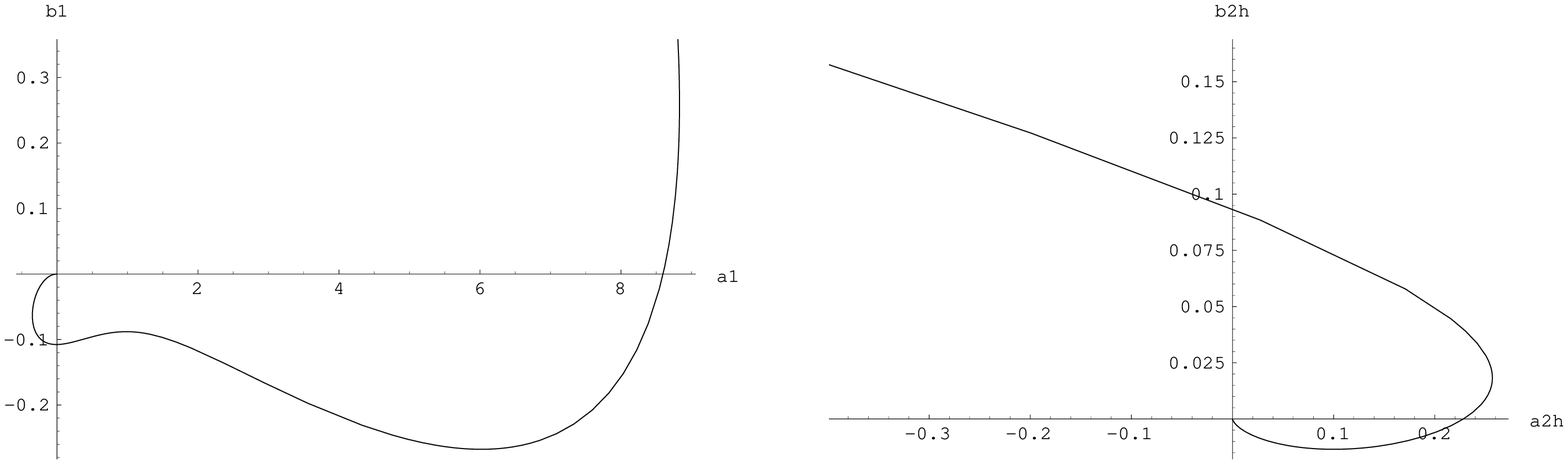}}
	\caption{Perturbation of the ground state for $\kappa=0.03$}
	\label{sg6}
\end{figure}
\begin{figure}[thbp]
	\epsfxsize=15cm
	\centerline{\epsfbox{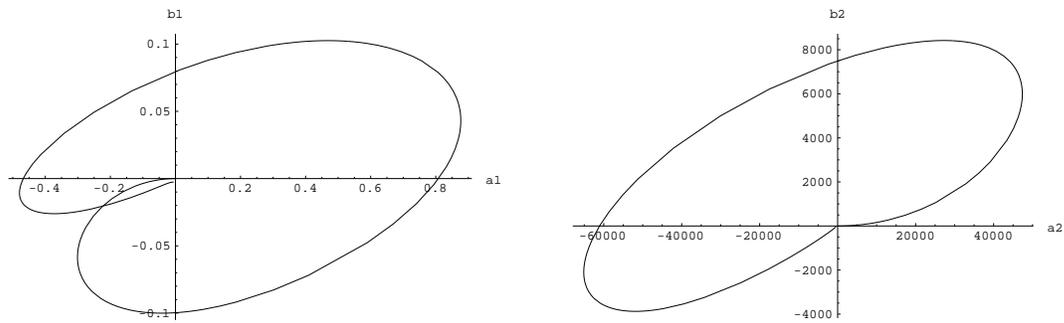}}
	\caption{Perturbation of the first excited state for $\kappa=10^{-6}$}
	\label{stab3}
\end{figure}
\begin{figure}[thbp]
	\epsfxsize=15cm
	\centerline{\epsfbox{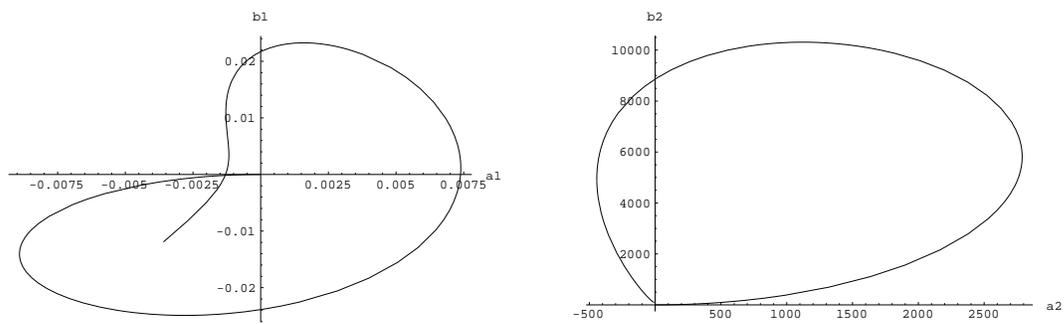}}
	\caption{Perturbation of the lowest negative-mass state for $\kappa=10^{-6}$}
	\label{stab4}
\end{figure}
\begin{figure}[tb]
	\epsfxsize=10cm
	\centerline{\epsfbox{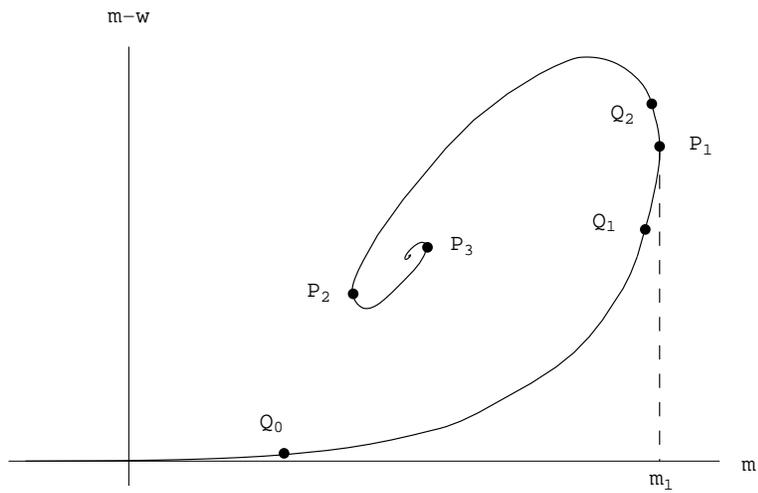}}
	\caption{Mass-Energy spectrum for the $n^{\mbox{\scriptsize{th}}}$ 
	excited state}
	\label{spiral}
\end{figure}

\begin{tabular}{ll}
\\
Mathematics Department, & Mathematics Department,\\
Harvard University, & The University of Michigan,\\
Cambridge, MA 02138  \hspace*{.5cm}(FF \& STY)\hspace*{1cm}
& Ann Arbor, MI 48109 \hspace*{.5cm} (JS)\\
\\
\end{tabular}

\begin{tabular}{ll}
email:&felix@math.harvard.edu\\
&yau@math.harvard.edu\\
&smoller@umich.edu
\end{tabular} 

\end{document}